\documentclass[useAMS, usenatbib, apj]{emulateapj}
\usepackage{afterpage}
\usepackage{tensor}
\usepackage{graphicx}
\usepackage{subfigure}
\usepackage{amsmath, amssymb, amsfonts}
\usepackage[normalem]{ulem}
\usepackage{multirow}
\usepackage{ulem}

\newcommand \rin {r_{\rm in}}
\newcommand \rout {r_{\rm out}}

\newcommand \divE {\nabla\cdot\vec{E}}

\newcommand \curlE {\nabla \times \vec{E}}
\newcommand \curlB {\nabla \times \vec{B}}
\newcommand \E {\vec{E}}
\newcommand \B {\vec{B}}
\newcommand \J {\vec{J}}
\newcommand \rlc {R_{\rm LC}}

\newcommand \dd {\partial}
\newcommand \beq {\begin{equation}}
\newcommand \eeq {\end{equation}}

\newcommand \lp {\left(}
\newcommand \rp {\right)}
\newcommand \lb {\left\{}
\newcommand \rb {\right\}}
\newcommand \ls {\left[}
\newcommand \rs {\right]}

\newcommand \rstar {r_{\star}}
\newcommand \thpc {\theta_{\rm pc}}
\newcommand \throt {\theta_{\rm rot}}

\newcommand \upc {u_{\rm pc}}
\newcommand \utw {u_{\rm pc}}

\newcommand \urot {u_{\rm rot}}

\newcommand \psicrit {\psi_{\rm crit}}
\newcommand \psirec {\psi_{\rm rec}}

\renewcommand \S {Section}

\let\oldhat\hat
\renewcommand{\vec}[1]{\boldsymbol{#1}}
\renewcommand{\hat}[1]{\oldhat{\mathbf{#1}}}

\newbox\grsign \setbox\grsign=\hbox{$>$} \newdimen\grdimen \grdimen=\ht\grsign
\newbox\simlessbox \newbox\simgreatbox \newbox\simpropbox
\setbox\simgreatbox=\hbox{\raise.5ex\hbox{$>$}\llap
     {\lower.5ex\hbox{$\sim$}}}\ht1=\grdimen\dp1=0pt
\setbox\simlessbox=\hbox{\raise.5ex\hbox{$<$}\llap
     {\lower.5ex\hbox{$\sim$}}}\ht2=\grdimen\dp2=0pt
\setbox\simpropbox=\hbox{\raise.5ex\hbox{$\propto$}\llap
     {\lower.5ex\hbox{$\sim$}}}\ht2=\grdimen\dp2=0pt

\shorttitle{}
\shortauthors{K. Parfrey, A. M. Beloborodov, and L. Hui}

\begin{document}

\title{Dynamics of Strongly Twisted Relativistic Magnetospheres}
\author{Kyle Parfrey\altaffilmark{1,2}, Andrei M. Beloborodov\altaffilmark{3}, and Lam Hui\altaffilmark{3,4}}
\affil{
$^{1}$Astronomy Department, Columbia University, 550 West 120th Street, New York, NY 10027, USA\\
$^{2}$Department of Astrophysical Sciences, Princeton University, Peyton Hall, Princeton, NJ 08544, USA; parfrey@astro.princeton.edu\\
$^{3}$Physics Department and Columbia Astrophysics Laboratory, Columbia University, 538 West 120th Street, New York, NY 10027, USA\\
$^{4}$Institute for Strings, Cosmology, and Astroparticle Physics (ISCAP), Columbia University, New York, NY 10027, USA
}

\label{firstpage}
\begin{abstract}
Magnetar magnetospheres are believed to be strongly twisted, due to shearing of the stellar crust by internal magnetic stresses. We present time-dependent axisymmetric simulations showing in detail the evolution of relativistic force-free magnetospheres subjected to slow twisting through large angles. When the twist amplitude is small, the magnetosphere moves quasi-statically through a sequence of equilibria of increasing free energy. At some twist amplitude the magnetosphere becomes tearing-mode unstable to forming a resistive current sheet, initiating large-scale magnetic reconnection in which a significant fraction of the magnetic free energy can be dissipated. This ``critical'' twist angle is insensitive to the resistive length scale. Rapid shearing temporarily stabilizes the magnetosphere beyond the critical angle, allowing the magnetosphere of a rapidly differentially rotating star to store, and dissipate, more free energy. In addition to these effects, shearing the surface of a rotating star increases the spindown torque applied to the star. If shearing is much slower than rotation, the resulting spikes in spindown rate can occur on timescales anywhere from the long twisting timescale to the stellar spin period or shorter, depending both on the stellar shear distribution and the existing distribution of magnetospheric twists.  A model in which energy is stored in the magnetosphere, and released by a magnetospheric instability, therefore predicts large changes in the measured spindown rate before SGR giant flares.
\end{abstract}


\section{Introduction}

Force-free magnetic fields, embedded in low-pressure perfectly conducting plasma, can store energy when the field is twisted into a non-potential state. This deformation can be the result of smooth shearing of a boundary surface, such as a heavy plasma or a conducting solid, in which the field's footpoints are frozen. The sudden release of the stored energy is a compelling model for solar flares and coronal mass ejections: energy is accumulated in the coronal force-free field due to motion of the heavy photospheric plasma, and liberated following a loss of magnetic equilibrium. The dynamic behavior of the solar corona is appropriately studied using non-relativistic magnetohydrodynamics (MHD), as there the Alfv\'{e}n speed is much smaller than the speed of light \citep[e.g.][]{2007Sci...317.1192T}.

Magnetars, neutron stars with ultra-strong magnetic fields $B \gtrsim 10^{14}$ G, also release magnetic energy in sudden bursts and flares \citep[e.g.][]{2006csxs.book..547W,2008A&ARv..15..225M}. An evaluation of the ability of the magnetospheres of these objects to store magnetic energy, and the stability of the resulting field configurations to energy release on dynamic timescales, requires study of the twisting problem in the relativistic plasma limit,
\beq
\frac{B^2}{8\pi} \gg \rho c^2 + U \,,
\label{eq:bsqgtrrho}
\eeq
where $\rho$ and $U$ are the mass and internal energy densities. In the magnetospheres of compact objects the Alfv\'{e}n speed is very nearly the speed of light, and the Lorentz force density vanishes even during rapid dynamic motion. Here, the evolution of the magnetosphere is described by force-free electrodynamics (also known as relativistic force-free MHD or magnetodynamics). The force-free approximation is well satisfied throughout the magnetosphere, except in certain regions (current sheets) where magnetic energy is dissipated.

When the magnetosphere is axisymmetric, the ``twist'' $\psi$ can be measured as the azimuthal angular displacement between a field line's two footpoints due to their relative shearing motion; $\psi = 0$ on all lines in the potential state. The study of current-carrying magnetar field configurations has largely been restricted to either the linear, weakly sheared limit $\psi \lesssim 1$ \citep{2009ApJ...703.1044B}, or the case of self-similar fields \citep{2002ApJ...574..332T, 2009MNRAS.395..753P}. While a self-similar model allows one to generate formal magnetospheric solutions with different amounts of shear, they cannot be connected as a realistic evolutionary sequence because this sequence would require compressive motions of the footpoints of magnetic field lines, and thus compression of the crust into which the field lines are frozen; this is forbidden as a neutron star's crust is nearly incompressible. Another tool is relaxation to a stationary twisted configuration satisfying prescribed boundary conditions on the star \citep{1986ApJ...309..383Y, 1994ApJ...423..847R}, which has recently been applied to the magnetar problem \citep{2011A&A...533A.125V}. 

Additionally, magnetar magnetospheres are not globally twisted---the shear is expected to be confined to a fraction of the stellar surface \citep{2009ApJ...703.1044B}. Further progress in magnetar theory requires the development of a fully nonlinear model for twists in realistic geometries, their stored energy, and the dynamical release of this energy when the field lines are overtwisted and force-free equilibrium is lost. One would also like to know the effect of the twists on the spindown rate of the star.

\subsection{Sequence of equilibria \label{sec:intro-seq-equilibria}}
As long as stellar rotation is neglected and purely static (equilibrium) configurations are considered, relativistic and non-relativistic force-free magnetospheres are described by the same equations. These equilibria have been extensively studied in the context of the solar corona. An equilibrium force-free configuration satisfies
\beq
\lp\curlB\rp \times \B = 0,
\label{eq:curlBcrossB}
\eeq
and so the magnetic field and steady-state current are parallel,
\beq
\curlB = \alpha(\mathcal{L})\B,
\eeq
where $\alpha$ is constant on each field line, labeled by $\mathcal{L}$. The boundary value problem posed by Equation~(\ref{eq:curlBcrossB}) can be investigated by setting $\alpha = \lambda\, g(\mathcal{L})$, where $\lambda$ is a constant and $g(\mathcal{L})$ is some prescribed function that is constant along field lines\footnote{
In axial symmetry, the poloidal field can be written in terms of one vector potential component $A_\phi$ as $\B_{\rm p} = \nabla A_\phi\times \hat{\vec{\phi}}/r\sin\theta$. (This $A_\phi = f/2\pi$, where $f$ is the poloidal flux function defined in Equation~(\ref{eq:definef}).) Equation~(\ref{eq:curlBcrossB}) is satisfied if the toroidal field has the form $B_{\phi} = F(A_\phi)/r\sin\theta$; then $\alpha = \rm{d}F/\rm{d}A_\phi$. A popular method of constructing a sequence of equilibria is by imposing a power-law relation between $F$ and $A_\phi$, $F = \lambda A_\phi^\gamma$ \citep{1971Ap&SS..14..464R,1976MNRAS.174..307M}.
}. 
A sequence of equilibrium states can be constructed by increasing $\lambda$ from zero (the potential field). Under quite general conditions, it can be shown using a virial theorem approach that there is a limiting value $\lambda^*$, such that there are no equilibrium solutions for $\lambda > \lambda^*$ \citep{1984ApJ...283..349A}. It was argued that reaching this limit implied the formation of current sheets \citep{1976MNRAS.174..307M} and the onset of dynamic behavior, which could trigger a solar flare \citep{1977ApJ...212..234L}.

Using virial relations, it can also be proven that there is a maximum energy associated with the sequence of force-free fields with a fixed normal magnetic field distribution on the stellar surface \citep{1984ApJ...283..349A}, and that the field lines become open to infinity for this maximum energy configuration \citep{1991ApJ...375L..61A,1991ApJ...380..655S}. For a twisted dipole configuration, the related asymptotic open field is a split-monopole-like state having the same flux distribution over the stellar surface as the original (untwisted) dipole field. Evaluating the energy of twisted configurations is useful as it shows how much energy can be stored in the non-potential magnetosphere; it also helps identify possible spontaneous transitions between twisted states. All field lines cannot spontaneously open to infinity, since the fully open field has the maximum energy. However, it is possible for a closed magnetic field to have higher free energy than a partially open field which is accessible by ideal MHD plasma motions \citep{1992ApJ...391..353W}. Therefore spontaneous partial opening is energetically allowed---the closed field can be unstable in an absolute sense.

A more physical way of constructing twisted equilibria is by specifying the shear, or angular separation between footpoints, of each field line; then a sequence of solutions can be generated by monotonically increasing the shear \citep{1978ApJ...224..668L}. Shearing the field line footpoints increases the magnetic pressure and energy of the field, causing the field lines to expand outward. As the field lines expand, a current layer forms, which becomes thinner and whose current density grows as the open-field configuration is approached. Following this sequence, one finds that the field lines open to infinity and that the layer becomes a current sheet, infinitely thin and with infinite current density, implying a tangential discontinuity in the magnetic field \citep{1989ApJ...344..471W, 1995ApJ...443..810W}. In the presence of any non-zero resistivity, these current layers would eventually be subject to reconnection, triggering dynamic motion and the release of stored energy in such an ``overtwisted'' magnetosphere.

The critical-point behavior of overtwisted magnetospheres has been investigated using several techniques. The $\alpha$-specified equilibrium sequence ends at a maximum $\lambda^*$, as described above. Using a perturbative expansion around the dipole potential field outside a magnetized sphere, \citet{1986ApJ...307..205L} showed that some smooth shearing profiles result in unbounded plasma displacements far from the sphere even for infinitesimally small shear. The field line expansion accelerates dramatically above a certain shear, as seen in studies using the magnetofrictional method \citep{1994ApJ...423..847R} and a self-similar model \citep{1995ApJ...443..810W}. \citet{2002ApJ...574.1011U} argues from principles of magnetostatic balance that field configurations, with footpoints frozen into spheres, will open to infinity at a finite shear (i.e. that no equilibrium solutions with closed field lines exist, where the field lines' footpoints have angular separation larger than a critical value). This is equivalent to saying that a finite shear suffices to take the energy of the field configuration to its theoretical maximum.

\subsection{Dynamics of overtwisted configurations \label{sec:intro-dynamics}}

The shearing problem has also been studied using time-dependent numerical MHD simulations, with parameters suitable for the solar corona. Simulations have the advantage of naturally testing the stability of each equilibrium state. Resistive simulations in a Cartesian geometry show that strongly sheared fields eventually form current sheets, undergo reconnection, and eject large plasmoids \citep{1988ApJ...328..830M, 1989SoPh..120...49B, 1992ApJ...393..800F}. Differential azimuthal footpoint motion of a dipole field, in opposite directions above and below the equator in a spherical geometry, eventually leads to accelerating field line expansion and unstable eruption \citep{1991ApJ...382..677S}.

The question of critical shear was investigated in detail by \citet[henceforth ML94]{1994ApJ...430..898M}. Using ideal MHD simulations, they showed that smoothly sheared field configurations first move, in a quasi-steady manner, through a series of quasi-equilibrium states. One possible scenario is that at a certain value of the applied shear the field enters a state of ideal magnetic non-equilibrium and the field lines open to infinity, eventually forming a tangential discontinuity in the magnetic field (current sheet). In their simulations, no eruptive behavior occurs during ideal evolution---after entering the non-equilibrium state the field is conjectured to transition smoothly to the (partially) open configuration (although this behavior is not observed, possibly due to the long equilibration timescales of inflated field lines). They do, however, find a critical shear separating distinct responses to resistivity: before the critical state is reached, the field slowly relaxes if resistivity is introduced, while after this point the introduction of a resistive term leads to reconnection in the current layer and the expulsion of a plasmoid.

\subsection{This paper}

In this paper, we study the dynamics of twisted relativistic magnetospheres (Equation~(\ref{eq:bsqgtrrho})) using time-dependent numerical simulations with the code \mbox{\textsc{PHAEDRA}} \citep{2012MNRAS.423.1416P}. The formulation of the problem is given in \S~\ref{sec:problem}, where we write down the equations of force-free electrodynamics, and describe the initial and boundary conditions. The numerical method is briefly described in \S~\ref{sec:numerics}. In Section~\ref{sec:equil}, we concentrate on non-rotating stars, and consider the sequence of nonlinear equilibria that is obtained by gradual shearing of the magnetic footpoints, the resulting inflation of poloidal field lines, and the free energy stored in the twisted magnetosphere. In \S~\ref{sec:dynamics} we describe the dynamic phase, entered when the magnetosphere becomes ``overtwisted'' and loses equilibrium. We address the critical twist amplitude at which equilibrium is lost and its dependence on the shearing rate and profile, the reconnection rate following the formation of the current sheet, and the relative amounts of energy expelled and dissipated. We then describe the evolution of the strongly twisted magnetospheres of stars in solid-body rotation in \S~\ref{sec:rotate}, where we discuss how rotation affects the equilibrium solutions, the critical twist amplitude, and the dynamic phase leading to reconnection. In particular, we describe how twisting significantly changes the spindown torque applied by the magnetosphere to the star. We also describe the small kicks delivered to the star by Alfv\'{e}n waves created during reconnection. Finally, we summarize and discuss our results in \S~\ref{sec:discussion}.

\section{Problem formulation}
\label{sec:problem}

\subsection{ Force-free electrodynamics}
\label{sec:ffe}

In the force-free (vanishing-inertia) limit, the MHD equations can be recast as Maxwell's equations with a nonlinear current density $\J$:
\beq
\dd_t \B = - \curlE , \qquad   \dd_t \E = \curlB - 4\pi\,\J,
\label{eq:maxwell}
\eeq
\beq
\J = \frac{\vec{B} \cdot \curlB - \vec{E} \cdot \curlE }{4\pi B^2} \B + \lp \frac{\divE}{4\pi}\rp \, \frac{\vec{E}\times\vec{B}}{B^2}\,.
\label{eq:current}
\eeq
The plasma response is contained within the current term, which is a unique function of the magnetic and electric fields and their derivatives. Evolved with these equations, the field configuration has identically vanishing Lorentz force density, $\rho_{\rm e} \E + \J\times\B=0$, where $\rho_{\rm e} = \divE/4\pi$ is the charge density. The electric field is perpendicular to the magnetic field (it is ``degenerate'') and to the current density (there is no Ohmic dissipation). Any configuration with a potential magnetic field (e.g.\ a dipole) and no electric field is a trivial solution of Equations~(\ref{eq:maxwell}) and (\ref{eq:current}).

\subsection{Model setup}
\label{sec:model}

The initial state in each simulation is a dipole magnetic field at rest, in a computational domain between the perfectly conducting stellar surface, at $r = \rstar$, and the outer boundary at $\rout$. This state has total energy $W_0$ in the computational domain,
\beq
W_0\lp \rstar, \rout\rp = \frac{\mu^2}{3}\lp \frac{1}{\rstar^3} - \frac{1}{\rout^3} \rp.
\label{eq:W0}
\eeq

We use spherical coordinates ($r, \,\theta, \,\phi$) with the polar axis aligned with the magnetic dipole axis. Field lines are labeled using the colatitude of the field-line footpoint $\theta_f$. We will also label field lines with the flux function $f$, which is the total magnetic flux through the stellar surface at colatitudes $\theta < \theta_f$. In the northern hemisphere ($\theta < \pi/2$) $f$ is given by
\beq
f(\theta_f) = 2\pi \rstar^2 \int_0^{\theta_f} B_r\lp \rstar, \theta\rp \sin\theta \,{\rm d}\theta.
\label{eq:definef}
\eeq 
For the surface normal field distribution of a dipole (which is unchanged by our axisymmetric azimuthal shearing) the flux function is 
\beq
f(\theta_f) = 2\pi \mu \frac{\sin^2\!\theta_f}{\rstar} = 2\pi \mu\frac{u\!\lp\theta_f\rp}{\rstar},
\eeq
where $u(\theta_f) = \sin^2\!\theta_f$ is the fractional flux function, $u(\theta_f) \equiv f(\theta_f)/f(\pi/2)$.

\subsection{Sources of twisting}

There are two sources of field line twisting in a magnetar's magnetosphere. The first is shearing of the stellar crust $\omega_c$, driven by internal magnetic stresses in the star. The surface may move gradually, or be subject to sudden motions (``starquakes'').  In either case motions are slow, in the sense that they develop over a timescale much longer than the light-crossing timescale of the inner magnetosphere. 

The second source of twisting is resistivity in the magnetosphere itself. It results in Ohmic dissipation $\E\cdot\J$ and field line twisting or untwisting at rate \citep{2011heep.conf..299B}
\beq
\frac{\dd\psi}{\dd t} = 2\pi c\, \frac{\dd\Phi}{\dd f}, 
\label{eq:resistive_twisting}
\eeq
where $\Phi$ is the voltage between the field line's footpoints, which is controlled by the threshold voltage for $e^\pm$ pair production, $\Phi \sim 10^9$ V \citep{2007ApJ...657..967B, 2009ApJ...703.1044B}. The resistive evolution according to Equation~(\ref{eq:resistive_twisting}) occurs on characteristic timescales of the order of years. The Ohmic dissipation tends to quickly erase electric currents on field lines closing near the star,  forming a potential cavity ($\nabla\times\B=0$), but the mechanism can actually increase the twist on more extended field lines where $\dd\Phi/\dd f > 0$. Reported shrinking hot spots on transient magnetars support the localization of twist on extended field lines \citep{2011heep.conf..299B}.

These two twisting mechanisms may operate concurrently, and their combined effect can be represented by an effective shearing rate $\omega$,
\beq
\frac{\dd\psi}{\dd t}(f) = \omega_c(f) + 2\pi c\, \frac{\dd\Phi}{\dd f} \equiv \omega(f).
\eeq
The resistive term, $2\pi c\, \dd\Phi/\dd f$, is negligible when the shearing rate is greater than 1~rad~yr$^{-1}$. In our simulations we use the effective rate $\omega$, attributing all of the twisting to surface shearing. This allows us to model the magnetosphere as being entirely ideal and force-free, except in thin current sheets where resistivity is expected.

\subsection{Boundary conditions \& shearing profiles}
\label{sec:profiles}
 
Our simulations concern stars whose magnetospheres are twisted by differential rotation of their surfaces. As the conducting stellar surface drags the frozen-in magnetic field, an electric field is induced in the static laboratory frame
\beq
\E = - \ls \lp\vec{\Omega}+\vec{\omega}\rp\times\vec{r} \rs \times \B\,,
\eeq
where we have decomposed the surface motion into solid-body rotation, $\vec{\Omega}$, and shearing, $\vec{\omega}$. The simulations are axisymmetric: $\vec{\Omega}$ and $\vec{\omega}$ are both always parallel to the magnetic axis.

\begin{figure*}[tp]
\centering
\includegraphics[width=165mm]{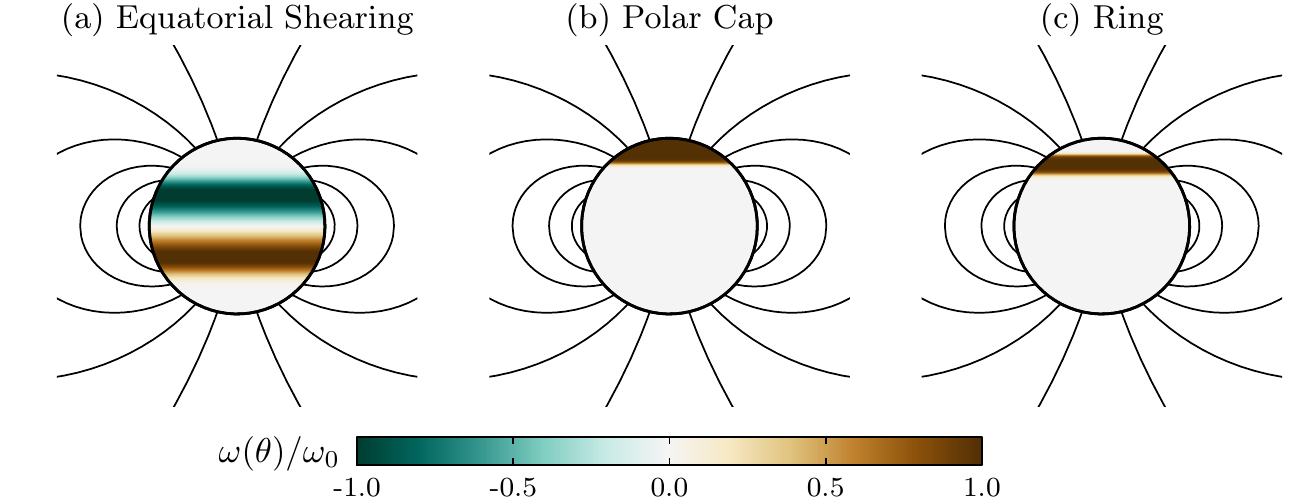}
\caption{The three shearing profiles used, normalized to the maximum shearing rate $\omega_0$. (a) Equatorial shearing, Equation~(\ref{eq:MLprof}), peaking at $\theta = \pi/2 \pm \pi/9$; (b) polar cap, Equation~(\ref{eq:PCprof}), extending down to $\thpc = 0.25\pi$; (c) ring profile, Equation~(\ref{eq:ringprof}), centered at $\theta_{\rm ctr}=0.25\pi$ and with width $\Delta = 0.05\pi$. The dipole field lines are equally spaced in $u$, between $u=0.05$ and 0.9, and $\kappa = 50$ in (b) and (c). \label{fig:profiles}}
\end{figure*}

In this paper we present results with three shearing profiles $\omega(\theta)$, which are shown in Figure~\ref{fig:profiles}. The first is identical to that used by ML94,
\beq
\omega_{\rm ML}(\theta) = \omega_0 \frac{\Theta}{\sin\theta}\exp\ls \lp 1 - \Theta^4\rp/4 \rs
\label{eq:MLprof}
\eeq
where $\Theta = (\theta - \pi/2)/\Delta\theta_{\rm m}$ and $\Delta\theta_{\rm m} = \pi/9$. This profile concentrates the shear near the equator, peaking at $\theta_\pm = \pi/2 \pm \Delta\theta_{\rm m}$. We label the applied shear by the azimuthal displacement $\psi$ between two footpoints located at $\theta_\pm$; all other field lines have twist amplitude smaller than $\psi$. Field lines above and below the equator are dragged in opposite directions, and so equatorial reflection symmetry is preserved.

In order to survey the dependence of solutions on shear profile, we also study two other models: the ``polar cap'' and ``ring.'' In the polar cap model, the twisting rate is  nearly constant from the magnetic axis down to some colatitude, and then decreases exponentially quickly to zero,
\beq
\omega_{\rm cap}(\theta) = \frac{\omega_{0}}{1 + \exp \ls \kappa \lp \theta - \thpc\rp \rs},
\label{eq:PCprof}
\eeq
where $\kappa$ determines the sharpness of the transition region whose center is at $\thpc$ (i.e. $\omega_{\rm cap}(\thpc) = \omega_0/2$).  The ring model twists a band of latitudes, centered at $\theta_{\rm ctr}$ and with angular half-width $\Delta$,
\beq
\omega_{\rm ring}(\theta) =  \frac{\omega_{0}}{1 + \exp \lb \kappa \ls \left| \theta - \theta_{\rm ctr} \right| - \Delta \rs \rb}.
\label{eq:ringprof}
\eeq
The ring extends in colatitude from $\theta_1 \equiv \theta_{\rm ctr} - \Delta$ to $\theta_2 \equiv \theta_{\rm ctr} + \Delta$. We generally set $\kappa \approx 50$. In both the polar cap and ring twisting profiles, only the northern ($\theta < \pi/2$) footpoints are moved, and the fields do not remain symmetric about the equator. We choose these models as simple one- and two-parameter families of shearing profiles, approximating step-function selection of twisted field lines.

In each simulation, the shearing rate $\omega_0(t)$ is smoothly increased from zero to its maximum value using a cosine bell. Depending on the problem, the shearing is either maintained at this constant rate until the end of the simulation, or, if a specific twist is to be implanted, the shearing rate is eventually smoothly returned to zero. 

A non-reflecting outer boundary condition is applied at $\rout$; its numerical implementation is described briefly in \S~\ref{sec:coordmaps}, and in more detail in \citet{2012MNRAS.423.1416P}.

\subsection{Units}

When describing the results of our simulations we will use the following units. Distance will be measured in units of $\rstar$, time in units of $\rstar/c$, angular velocity in units of $c/\rstar$, magnetic field in units of $\mu/\rstar^3$, and current density in units of $c\mu/\rstar^4$. All angles are measured in radians.

\section{Numerical method}
\label{sec:numerics}

We solve the equations of force-free electrodynamics, Equations~(\ref{eq:maxwell}) and (\ref{eq:current}), using the parallel pseudospectral simulation code \textsc{PHAEDRA}; see \citet{2012MNRAS.423.1416P} for a detailed description of the code. 

The code calculates spatial derivatives by expanding the fields along each coordinate direction in global orthogonal basis functions (Chebyshev polynomials in the radial direction, cosine and sine functions in the meridional), calculating the coefficients of the derivative series in spectral space, and transforming back to real space. The variables are temporally integrated at each point in real space using Runge-Kutta time stepping. This method converges quickly to smooth solutions with increasing grid resolution, and has low numerical diffusion and dissipation.

\subsection{Filtering \& resistivity}
\label{sec:resist}

Spectral filtering is applied at each time step to prevent aliasing instability and the accumulation of high-wavenumber noise from sharp features in the solution. If a function $u(x)$ is expanded in a set of basis functions $T_n(x)$, the filtered function, $\mathcal{F}u(x)$, is given by
\beq
\mathcal{F}u(x) = \sum_{n=0}^{N-1} \sigma\lp\frac{n}{N-1}\rp \tilde{u}_n T_n(x)\,,
\eeq
where $\tilde{u}_n$ are the discrete expansion coefficients and $\sigma(\eta)$ is the filter function,
\beq
\sigma(\eta) = \exp\lp-\alpha \eta^{2p}\rp.
\eeq
We use two filters: one of very high order to prevent aliasing instability ($2p = 36$, $\alpha = - \ln{\epsilon_M}$, where $\epsilon_M$ is machine precision) and one of eighth order to maintain stability in the presence of discontinuous current sheets ($2p = 8$, $\alpha = \alpha_{\rm SSV} =$ 0.005--0.1). The effect of this second ``super spectral viscosity'' filter can be interpreted as an eighth-order hyper-resistivity, and is the dominant source of dissipation in our simulations. Because it is of high order, this resistivity is negligible on all resolved scales, only acting when the field gradients approach the grid scale. This restricts the dissipation to regions with high current density, where it is physically expected. Equivalently, the filter sets the reconnection length scale to be very close to the grid scale. Our solutions can be viewed as being ideal up to the point of current sheet formation, at which point resistivity is introduced where it is required. Increasing resolution decreases the reconnection length scale, allowing current sheet formation to proceed further before the current layer becomes resistive.

\subsection{Coordinate maps}
\label{sec:coordmaps}

In simulations of non-rotating stars, where the initial field is a standard dipole, the closed field lines extend to arbitrary distances, and outgoing waves on all field lines should be able to return to the star. This can be problematic when performing calculations in a finite domain. (This problem does not arise in studies of rotating stars, because outside the light cylinder the field lines are open to infinity, and can be safely truncated.) In some cases in Sections \ref{sec:equil} and \ref{sec:dynamics}, we use an exponential coordinate mapping to place the computational boundary at sufficient distance that no waves reach it from the star over the length of the simulation; in longer simulations we include an absorbing layer near the outer boundary, placed far enough away that the removal of outgoing waves has little effect. In all simulations, a non-reflecting boundary treatment, based on an approximate characteristic decomposition of the equations of motion, is implemented at $r = \rout$. The exponential coordinate map we use is 
\begin{align}
r(x)  &=  \exp\left\{Q \ls g_{\mathrm{asin}}(x) + 1\rs \right\} \rin \;, \label{eq:expmap} \\
Q &= \frac{1}{2} \ln\left(\frac{2 + \zeta \Delta\theta}{2 - \zeta \Delta \theta}\right)(N_r-1) \;, \nonumber
\end{align}
giving $r \in [\rin,\rout]$ where $\rout = e^{2 Q}\rin$, and $g_{\mathrm{asin}}(x)$ is the arcsine map which reduces the clustering of the Chebyshev nodes, $x \in [-1,1]$, near the end points. If $g_{\mathrm{asin}}(x)$ were exactly equispaced the resulting grid would have spacings $\Delta r$ that lay on a line through the origin with slope $\zeta \Delta \theta$. Here $\Delta\theta = \pi/N_{\theta}$ is the average  angular grid spacing, and $N_r$ and $N_\theta$ are the number of grid points in the radial and meridional directions.

In simulations of non-rotating stars we also use a transformation in the meridional direction,
\beq
\tilde{\theta} = \theta + \frac{\gamma}{2}  \sin (2\theta)
\eeq
with $\gamma = 0.4$, to smoothly increase the grid resolution near the equator. With rotation, it is important to resolve well the narrow open flux bundles near the poles, and so we do not use this mapping.

\section{Equilibria}
\label{sec:equil}

\subsection{Reaching equilibrium}
\label{sec:reacheqlm}

In this section we discuss the equilibrium solutions of a dipole magnetosphere, part of whose flux has been twisted through an angle $\psi$. Our simulations begin with a dipole field with no twist, the field being everywhere poloidal and potential. Shearing of the stellar surface begins at $t=0$, launching current- and charge-carrying Alfv\'{e}n waves into the magnetosphere and causing the twisted field lines to acquire toroidal magnetic field $B_{\phi}$. An equilibrium solution at a specific $\psi$ can be found by shearing the surface up to that angle, halting the shear motion, and allowing the magnetosphere to relax to a static configuration over several wave-crossing times of the twisted field lines\footnote{The simulated magnetosphere is nearly ideal (as long as no current sheets form); therefore the resulting twisted configuration will have the same twist amplitude $\psi$ as imparted by the surface motion.}. Alternatively, if the shearing is slow enough the magnetosphere will smoothly move through a sequence of quasi-equilibrium states. These are close to the true equilibria, which are approached as the twisting rate is decreased; the deviation can be measured with a scalar virial equation, as described in \S~\ref{sec:energy}. In this paper, most ``equilibrium'' solutions described, including all where a quantity is shown smoothly varying with $\psi$, are these quasi-equilibria.

\begin{figure*}[tp]
\centering
\includegraphics[width=165mm]{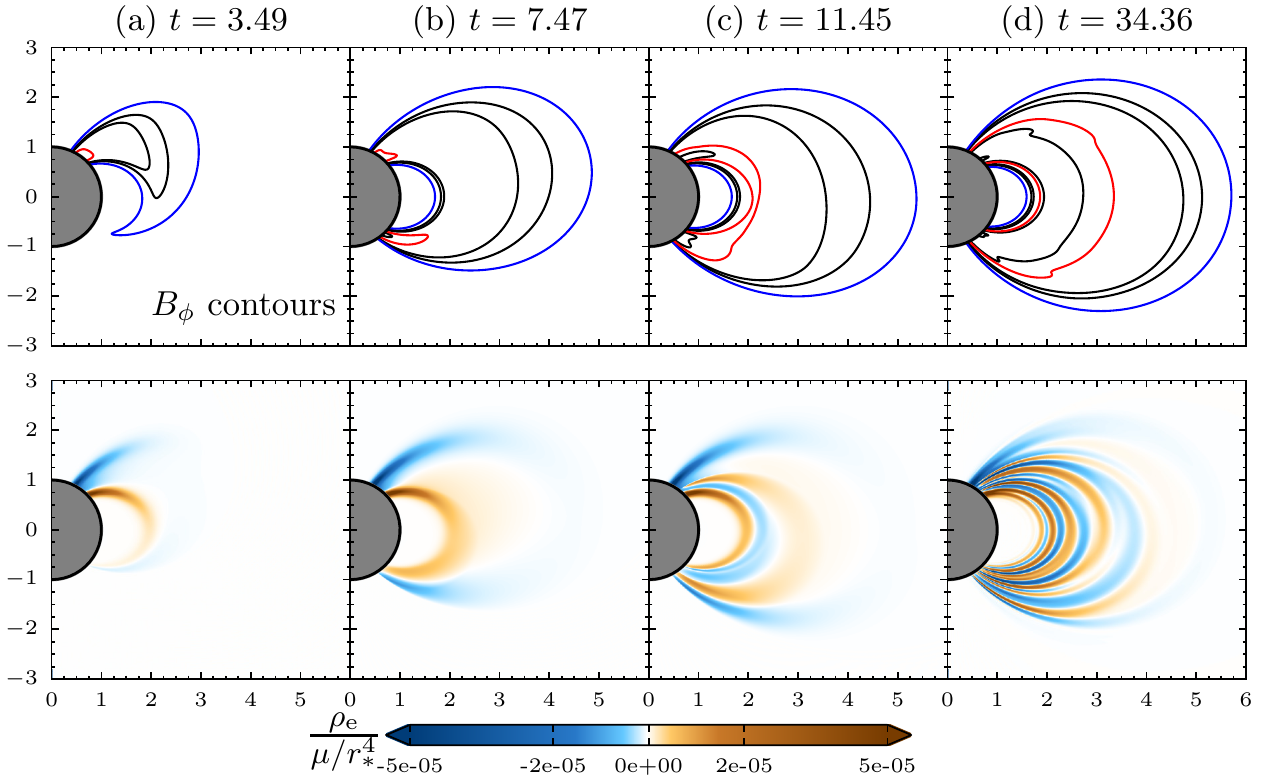}
\caption{Continuously twisted magnetosphere reaching quasi-equilibrium (initial transient phase). A ring extending from $\theta/\pi =$ 0.15 to 0.25 is smoothly brought from rest to $\omega_0 = 10^{-4}\, c/\rstar$ by $t=5\, \rstar/c$. Upper panels: contours of toroidal field at $-B_{\phi} = 10^{-6}$ (blue), $5\times 10^{-6}$, $10^{-5}$, $5\times 10^{-5}$ (red),  $10^{-4}$, and $5\times 10^{-4}\; \mu/\rstar^3$. Lower panels: charge density $\rho_e = \divE/4\pi$ in units of $\mu/\rstar^4$. Time is indicated in units of $\rstar/c$. \label{fig:wavetoeqlm}}
\end{figure*}

The evolution of an initially untwisted configuration to a twisted quasi-equilibrium state is shown in Figure~\ref{fig:wavetoeqlm}. A ring in the northern hemisphere is brought from rest to a constant twisting rate $\omega$ by $t = 5$, injecting an Alfv\'{e}n wave into the magnetosphere (Figure~\ref{fig:wavetoeqlm}a). The wavefront is sheared by the dipole field geometry, as the wave moves along diverging field lines. The wave reaches the field lines' southern footpoints and is reflected (Figure~\ref{fig:wavetoeqlm}b); subsequently, waves bounce backward and forward on the closed field lines, establishing quasi-equilibrium over the majority of the twisted flux by $t\approx 35$ (Figure~\ref{fig:wavetoeqlm}d), even as the ring continues to slowly rotate. Note that the quasi-equilibrium field is very close to being reflection-symmetric about the equator, even though only the northern footpoints are in motion, because the shearing timescale is much longer than the wave-crossing time scale on these field lines. This final state is very similar to that which is found if the twisting is immediately turned off and the configuration is allowed to relax to equilibrium (the wiggles in the $B_{\phi}$ contours are erased as the corresponding field lines equilibrate). This simulation was performed in a domain $1 \leq r \leq 60$ with grid size $N_r \times N_\theta = 512\times 375$.

One can think of the ideal (dissipationless) twisted configuration as a superposition of Alfv\'{e}n waves created by the polar-cap motion, which are trapped in the closed magnetosphere where they continue to bounce off the perfectly conducting surface. The waves are more easily seen in the charge density distribution (lower panels of Figure~\ref{fig:wavetoeqlm}). They are stretched to smaller perpendicular scales over time, as the background geometry increasingly shears each succeeding reflected wave. In a perfectly ideal plasma the waves would bounce backward and forward indefinitely, and be sheared ever thinner; in our numerical solutions, their thickness reaches the grid scale, and they are gradually smoothed and removed by the spectral filters. Eventually, the waves excited by the initial acceleration of the ring are damped, and the subsequent evolution is a slow progression through a sequence of configurations with increasing $\psi$, which are almost identical to exact equilibria.

\subsection{Sequence of equilibrium solutions}
\label{sec:seqeqlm}

Once the twisted field lines have reached quasi-equilibrium, the sequence of force-free equilibria can be investigated, by slowly shearing the stellar surface at a constant rate.  The numerical solution moves quasi-statically through the sequence as the accumulated twist $\psi$ on the field lines increases. 

For a perturbative expansion in small $\psi \ll 1$, the first-order departure from the potential field is the addition of a toroidal component; the second-order effect is a slow expansion of the poloidal field lines, as the additional magnetic pressure due to the toroidal field modifies the pressure--tension force balance, pushing the field lines outward \citep{1986ApJ...307..205L}.

\begin{figure*}[tp]
\centering
%\vspace{-14pt}
\includegraphics[width=160mm]{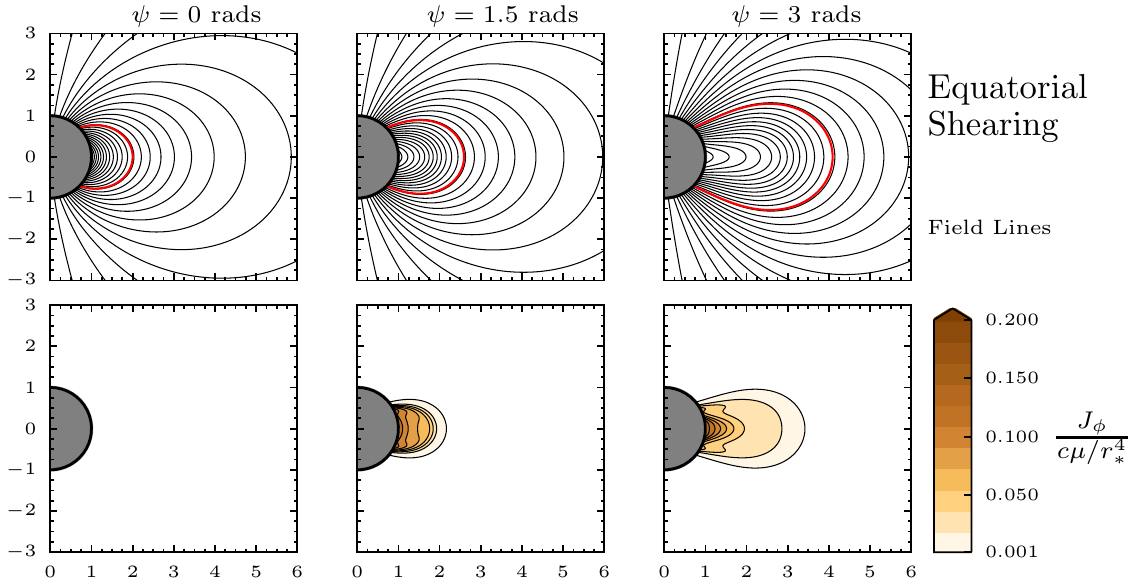}
\caption{Equilibrium solutions for the equatorial (ML94) shearing profile, Equation~(\ref{eq:MLprof}). Upper panels: poloidal field lines, equally spaced in flux function in the range $u =$ 0.01--0.975 with spacing $\Delta u \approx 0.04$, and a red field line at $u=0.5$. Lower panels: filled contours of $J_{\phi}$, equally spaced between $J_\phi =$ 0.001 and 0.2 $c\mu/\rstar^4$ with step $\Delta J_\phi \approx 0.018\,c\mu/\rstar^4$. Axes are labeled in units of $\rstar$. \label{fig:equil_solns_1}}
\end{figure*}

\begin{figure*}[tp]
\centering
%\vspace{-14pt}
\includegraphics[width=160mm]{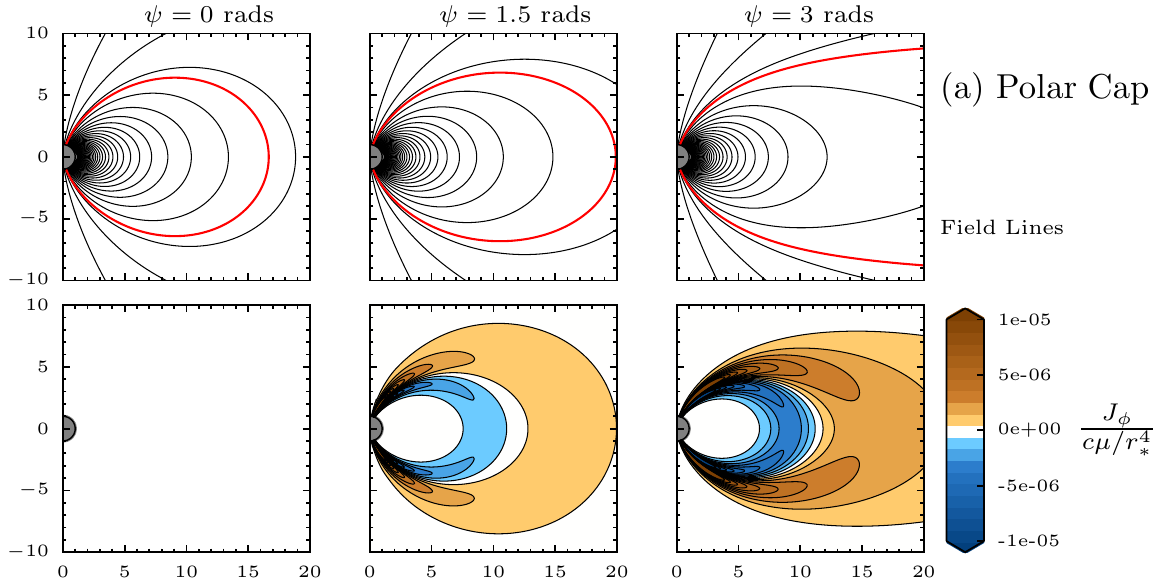} 
\includegraphics[width=160mm]{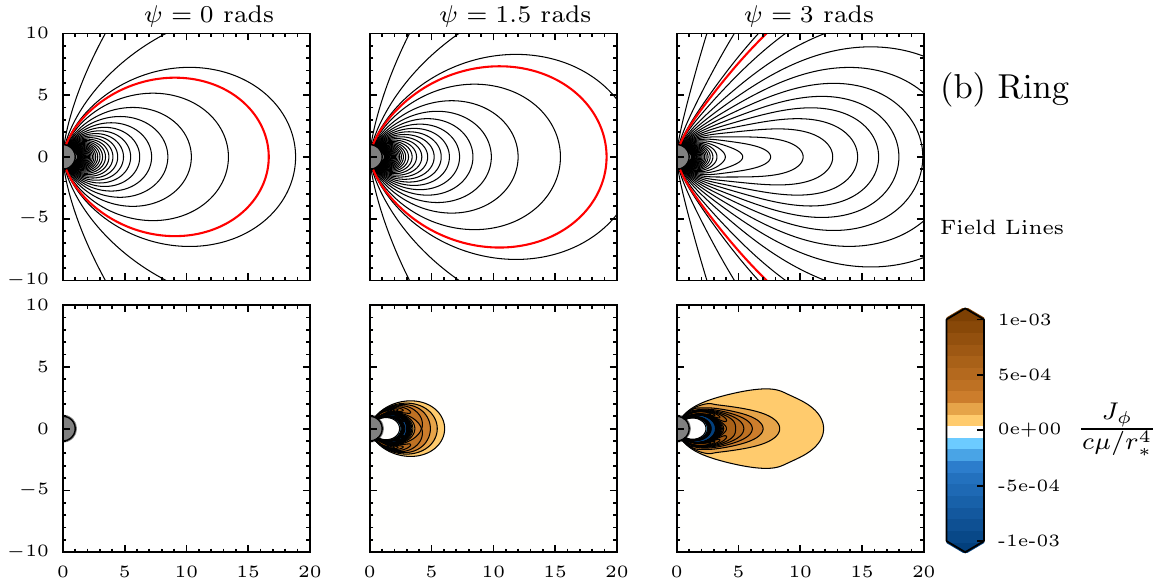} %\vspace{-12pt}
\caption{Equilibrium solutions for two shearing profiles: (a) polar cap $\lp \thpc = 0.1\pi\rp$ and (b) ring (bounded by $\theta_1 = 0.15\pi$ and $\theta_2 = 0.25\pi$). Upper panels: poloidal field lines, equally spaced in flux function in the range $u=$ 0.01--0.8 with spacing $\Delta u \approx 0.02$, and a red field line at $u=0.06$. Lower panels: 20 filled contours of $J_{\phi}$, equally spaced in the ranges shown. Axes are labeled in units of $\rstar$. \label{fig:equil_solns_2}}
\end{figure*}

Our numerical procedure allows us to generate equilibria for general surface shearing profiles and twist angles. Solutions at $\psi = $ 0, 1.5, and 3 for three shearing profiles are shown in Figure~\ref{fig:equil_solns_1} and Figure~\ref{fig:equil_solns_2}. These profiles were chosen to illustrate shearing confined to regions near the equator, at the pole, and in mid-latitudes (see Figure~\ref{fig:profiles}). The simulations of the equatorial shearing model described in this section were performed with a grid having $N_r\times N_\theta = 384\times 255$ and a computational domain $1 \leq r \leq 60$, while those of the polar cap and ring shearing models used a grid of $640\times 255$ points and a domain $1 \leq r \leq 2155$. The solutions shown are stable; if the shearing is arrested they remain static indefinitely, which we have explicitly verified over many magnetospheric light-crossing times. This also demonstrates that artificial numerical resistivity is practically absent in our pseudospectral code, as long as no current sheets form.

In all three cases it can be seen that there is noticeably more poloidal expansion between $\psi =$ 1.5 and 3 than between $\psi =$ 0 and 1.5. The toroidal current density, $J_{\phi}$, is everywhere positive for the equatorial shearing model. In contrast, for the polar cap and ring models, $J_\phi$ changes sign as one moves away from the pole toward the region where the shearing rate is zero. The early stages of the formation of the equatorial current sheet are evident in Figures~\ref{fig:equil_solns_1} and \ref{fig:equil_solns_2}b, including the concentration of toroidal current near the equator. Equatorial shearing causes all the field lines to expand significantly; eventually (in the absence of resistivity) every field line will open to infinity given sufficient twisting, and the magnetosphere will be brought to the maximum-energy configuration for a dipole surface flux distribution. The ring shearing profile, on the other hand, displays a clear distinction between twisted, inflating, field lines and the untwisted field below; the final state in this sequence will be a partially open field. The polar cap solution (Figure~\ref{fig:equil_solns_2}a) shows less progress toward forming the current sheet by $\psi =$ 3; as with ring shearing, the final state at large $\psi$ will also be partially open.

The equatorial shearing solutions are similar to those found by ML94, using the same shearing profile but evolving the time-dependent equations of non-relativistic MHD in a restricted ``zero (plasma) beta'' formulation. This similarity is expected, because the force-free equilibrium states, as determined by Equation~\ref{eq:curlBcrossB}, are independent of the plasma motions by which they are produced.
 
\begin{figure*}[tp]
\begin{center}
\includegraphics[width=160mm]{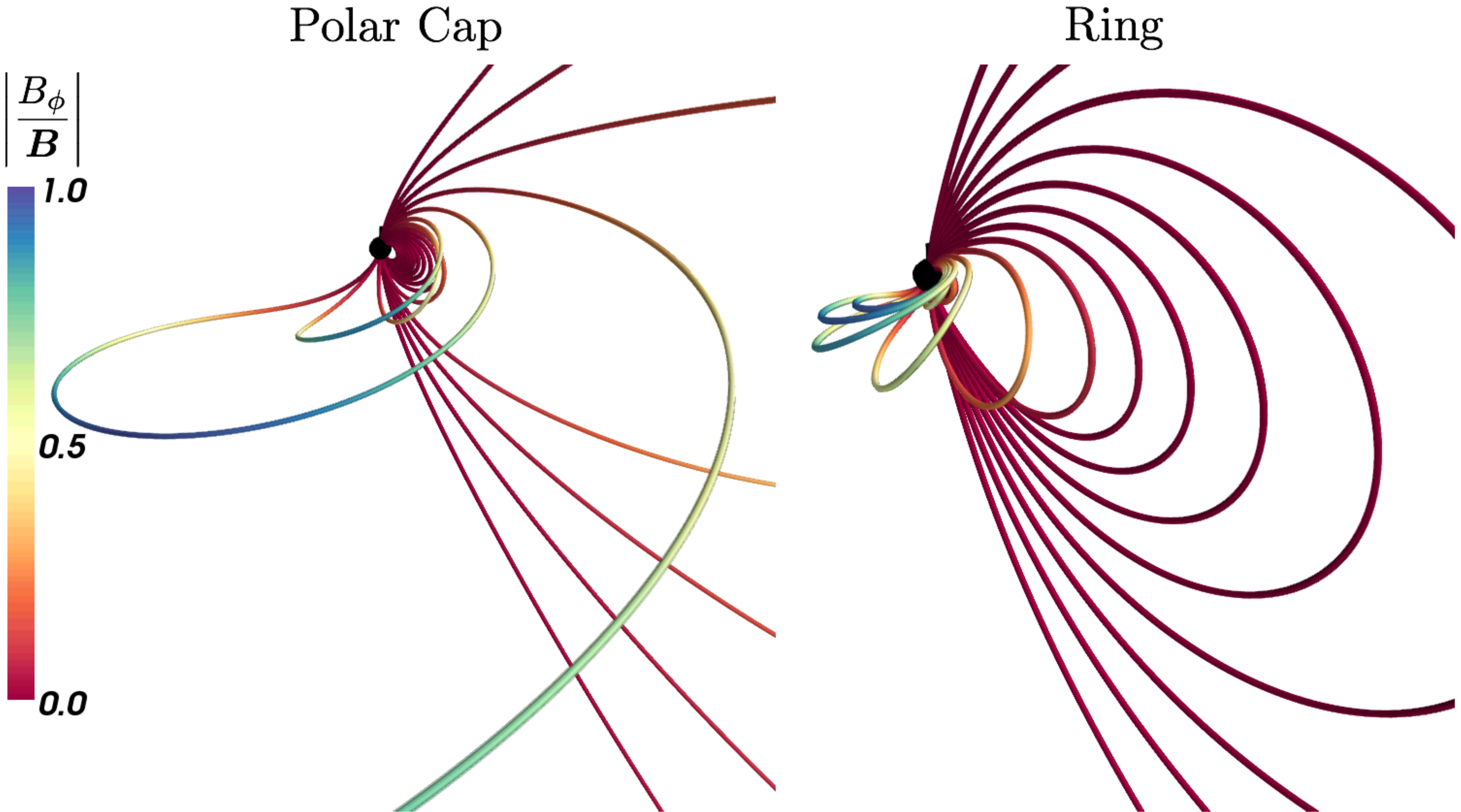}
\caption{ Three-dimensional field lines, for the same polar cap and ring shearing models as in Figure~\ref{fig:equil_solns_2}, at $\psi =$ 3. The lines are colored by the fractional contribution of the toroidal field, $\left|B_{\phi}/\vec{B}\right|$, at each point along their lengths. 20 lines are drawn from each hemisphere, equally spaced in colatitude between $\theta/\pi =$ 0.04 and 0.25. \label{fig:3dlines}}
\end{center}
\end{figure*}

%% 3D field lines
Field lines, drawn in three-dimensional space, are shown in Figure~\ref{fig:3dlines} for the polar cap and ring shearing models used in Figure~\ref{fig:equil_solns_2}, again at $\psi =$ 3. At this large twist angle the toroidal magnetic field becomes dominant near the equator and small away from it. This means that field lines which are pushed away from the equatorial region by the expansion, including those attached to the twisted polar cap, become predominantly radial near the star (and out to increasing distances from it as $\psi$ grows).

The expansion of the field lines occurs in two phases. At small twist amplitude, the poloidal field is only weakly affected by shearing, and the maximum height of each field line above the stellar surface, $R_{\rm max}$, increases slowly with $\psi$. As one steps through the sequence of quasi-equilibria, the magnetosphere eventually becomes much more sensitive to $\psi$, and $R_{\rm max}$ then increases rapidly with twist angle.

\begin{figure}[tp]
\begin{center}
\includegraphics[width=85mm]{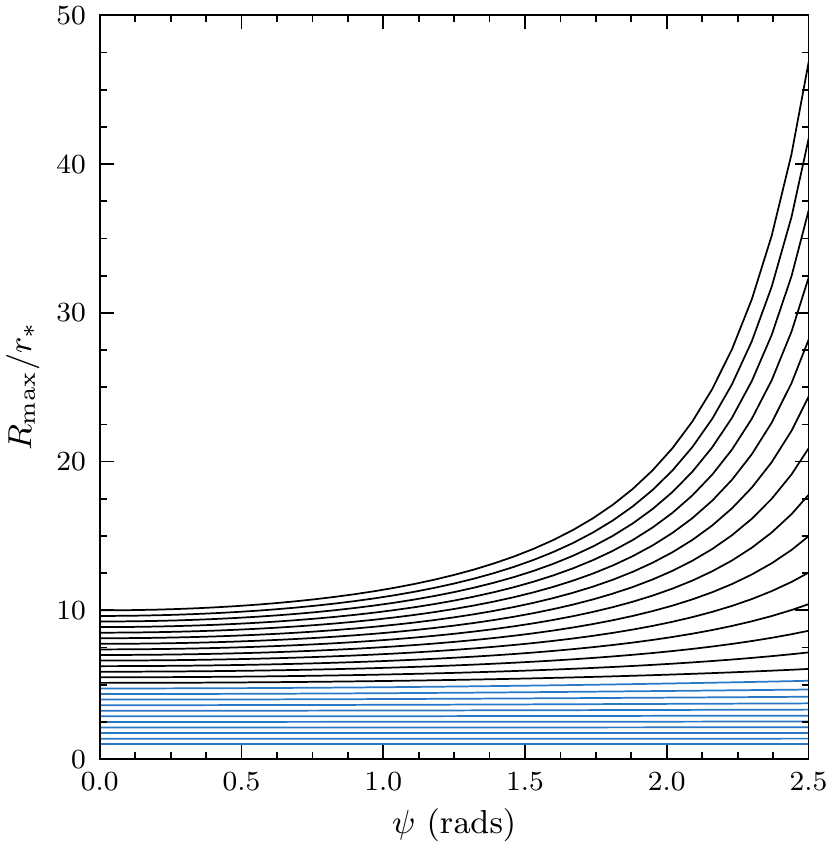}
\caption{$R_{\rm max}$ versus twist angle, where $R_{\rm max}$ is the maximum height of each field line, for a twisted polar cap with $\thpc = 0.15\, \pi$. The blue curves represent untwisted field lines, attached to the star outside the polar cap. \label{fig:rmax}}
\end{center}
\end{figure}

When the magnetosphere enters the second, fast-expansion phase depends on the profile of the applied shear, and in particular on whether the twisting is applied all the way to the pole (polar cap) or if there are untwisted field lines surrounding the sheared flux (ring profile); in the latter case the untwisted lines help to contain the twisted flux, and entry into the rapidly inflating regime is delayed. In Figure~\ref{fig:rmax} the two phases are illustrated for a polar cap shearing model. In this case, the magnetosphere inflates above a twist of $\psi \sim 1.75$. The lowest-lying field lines are in the untwisted zone inside the sheared flux, and do not participate in the expansion. During rapid inflation the entire magnetosphere is no longer in quasi-equilibrium---the outermost field lines expand dynamically at the speed of light. However, for twist amplitudes $\psi \lesssim 3$ the magnetosphere would gently relax to a nearby equilibrium state, without change of field connectivity, if the surface shearing stops. The behavior at larger twist amplitudes will be discussed in \S~\ref{sec:dynamics}.

\subsection{Energy of equilibria}
\label{sec:energy}

The minimum energy state for a magnetosphere with a given distribution of $B_r$ on the stellar surface is the potential $\lp\curlB=0\rp$ field that has only closed field lines. Shearing does work against the field lines' tension, and transfers energy from the star to the magnetosphere. At small twist angle $\psi \ll 1$, this increase is due to the $B_{\phi}^2$ contribution added to the magnetic energy density, while the poloidal field is hardly changed. As $\psi$ increases and the field lines expand, the additional energy is increasingly stored in the poloidal field components. The limiting maximum energy configuration of sheared fields is the fully open field, in which all field lines extend to infinity and the toroidal component is everywhere zero \citep{1991ApJ...375L..61A,1991ApJ...380..655S}. For a dipole potential field in an infinite domain ($\rout\rightarrow\infty$) having energy $W_0$, this fully open state has energy $W_{\rm open} = 1.662\, W_0$ (\citealt{1972ApJ...174..659B}; ML94).

The total magnetic energy, $W$, can be found by integrating $B^2$ over volume,
\beq
W \equiv \frac{1}{8\pi}\int_V  B^2 {\rm d}V,
\eeq
where $V$ is the volume of the computational domain (excluding the outer absorbing layer if present). The energy of any equilibrium force-free configuration outside a surface is related to the distribution of $\B$ on that surface by a scalar virial theorem \citep[e.g.][]{1984ApJ...283..349A}. The energy expected in the computational domain from the virial theorem, $W_{\rm vir}$, can then be found by subtracting the energy that, in equilibrium, should lie beyond its outer boundary:
\beq
W_{\rm vir} \equiv W_{\infty}(\rstar) - W_{\infty}(r_{\rm out}),
\label{eq:Wvir}
\eeq
where 
\beq
W_{\infty}(r) \equiv \frac{r^3}{4} \int_0^\pi \lp B_r^2 - B_{\theta}^2 - B_{\phi}^2 \rp \sin\theta\, {\rm d}\theta
\eeq
is the energy of an equilibrium state, integrated from $r$ to infinity. Since an equilibrium solution must have $W = W_{\rm vir}$, these quantities can be used to test how close our quasi-equilibrium solutions are to equilibrium. 

The twist free energy is defined by $W_{\rm tw} = W - W_0$ where $W_0$ is the untwisted dipole energy given by Equation~(\ref{eq:W0}). At small twist amplitude $\psi \ll 1$, the free energy is just the energy of the toroidal magnetic field, which is given by \citet{2009ApJ...703.1044B},
\beq
W_{\rm tw} \approx \int_{r>\rstar}\frac{B_\phi^2}{8\pi}\,{\rm d}V = \frac{\mu}{2 c \rstar} \int_0^1 I(u)\, \psi(u)\, {\rm d}u,
\label{eq:Wtw}
\eeq
where $I(u)$ is the poloidal current function (defined in the same way as the poloidal flux function $f$, except that one replaces $B_r$ with $J_r$ in Equation~(\ref{eq:definef})). By Stokes' theorem, the toroidal magnetic field and poloidal current function are related by $B_\phi = 2I/c r \sin\theta$, and integrating $B_\phi$ along a field line gives a relation between $I$ and $\psi$,
\beq
\psi = \frac{4 I \rstar^2}{u^2 c \mu} \sqrt{1 - u}.
\label{eq:psi_I}
\eeq 
In our polar cap models, $\psi(u) \approx$ constant for $0 \leq u \leq \upc$ and $\psi(u) = 0$ elsewhere; combining this profile, Equations~(\ref{eq:Wtw}) and (\ref{eq:psi_I}), and $W_0 = \mu^2/3\rstar^3$ one estimates the total energy of the twisted configuration to be
\beq
W \approx \lp 1 + \frac{\psi^2 \utw^3}{8} \rp W_0, \qquad \psi \lesssim 1.
\label{eq:Westimate}
\eeq
(Note that $\utw$ is not identical to the $u_*$ used in \citet{2009ApJ...703.1044B}, hence the difference in numerical coefficients.)

\begin{figure}
\begin{center}
\includegraphics[width=87mm]{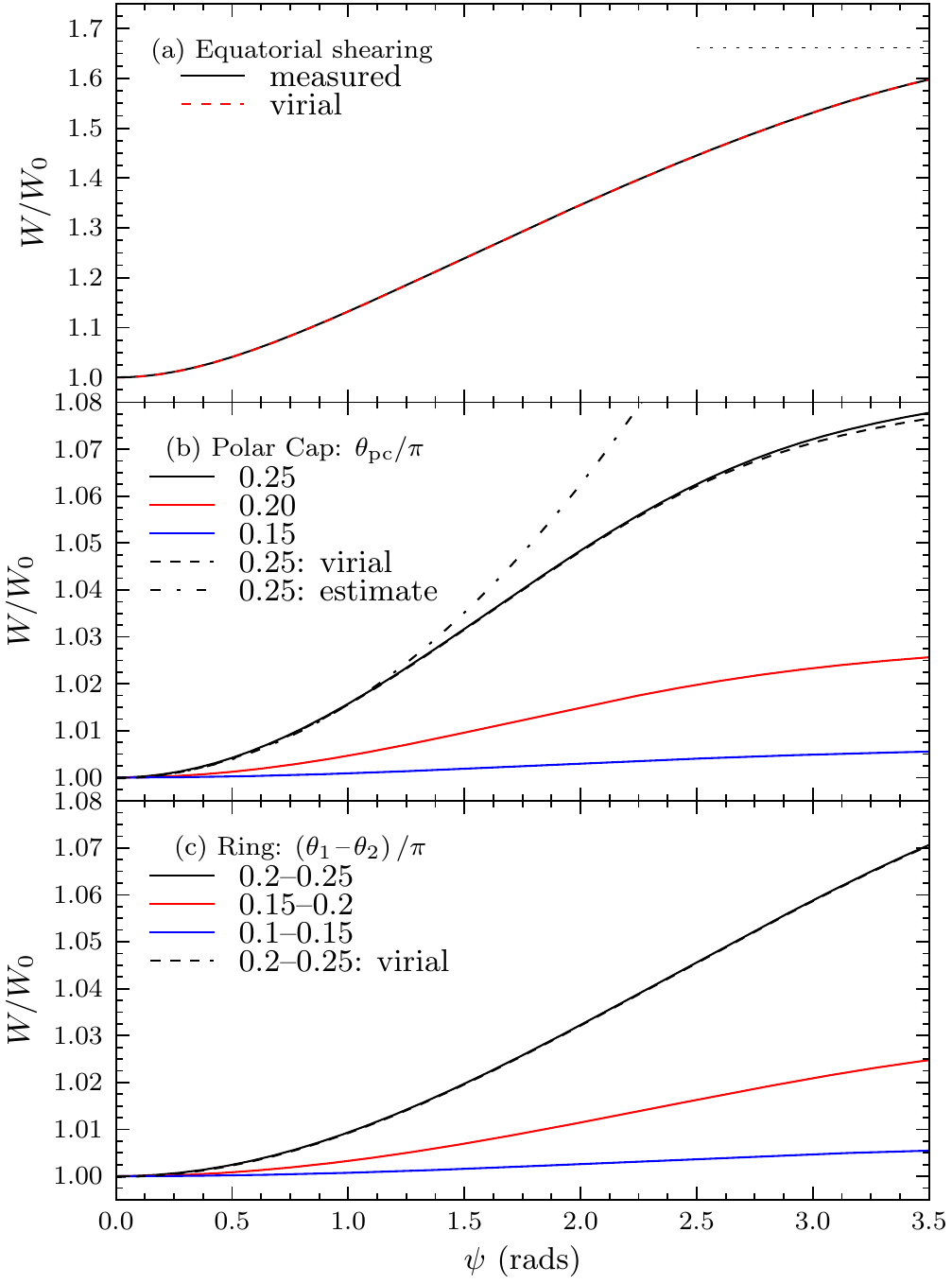}
\caption{Total energy on the grid, in units of the energy of the untwisted dipole $W_0$, for (a) equatorial, (b) polar cap, and (c) ring twisting profiles. The dashed lines show the value expected from the equilibrium force-free virial theorem, Equation~(\ref{eq:Wvir}), for the top curve in each panel. The dashed-dotted curve gives the analytic estimate, Equation~(\ref{eq:Westimate}), corresponding to the top polar cap model. The horizontal dotted line in (a) indicates the energy of the completely open field. \label{fig:energy_equil}}
\end{center}
\end{figure}

The measured energy $W$ as a function of $\psi$ is shown for several models in Figure~\ref{fig:energy_equil}, where it is also compared to $W_{\rm vir}$ (dashed lines) and the analytical estimate in Equation~(\ref{eq:Westimate}) (dashed-dotted line). For the equatorial shearing model (Figure~\ref{fig:energy_equil}a) the energy approaches, but does not exceed, the energy of the completely open field (dotted line), and ${\rm d}^2W/{\rm d}\psi^2 < 0$ near the open configuration. The solution is always very close to equilibrium ($W = W_{\rm vir}$) for the twist angles displayed. 

In panels (b) and (c) of Figure~\ref{fig:energy_equil} the energy is given for three polar cap models, and three ring models, of thickness $2\Delta = 0.05\pi$. Each polar cap model, of extent $\thpc$, corresponds to one ring model whose lower boundary $\theta_2$ is at the same colatitude $\thpc$. In each case, the energy of the twisted polar cap solution is greater than that of the corresponding ring solution, and the difference increases as the ring is selected closer to the equator. The analytical estimate, Equation~(\ref{eq:Westimate}), is a good approximation to the energy of the polar cap solution at twist angles $\psi \lesssim 1.25$, and overestimates the energy at larger twists. 

At twist angles above $\psi \sim 3$, the energy curve representing the top polar cap solution begins to depart slightly from the energy of an equilibrium state, indicating that the magnetosphere leaves the sequence of equilibria and enters a dynamical state. The ring solutions are still in equilibrium even at $\psi = 3.5$, because of the confining effect of unsheared field lines blanketing the twisted flux. We next turn our attention to what happens when the magnetosphere is subjected to large twist angles, and the quasi-equilibrium approximation ceases to be valid.

\section{Dynamics of overtwisted magnetospheres}
\label{sec:dynamics}

When a sufficiently large twist $\psi$ is applied, the magnetosphere must necessarily leave the sequence of force-free equilibria and enter a fully dynamical state. This has been argued in two ways in previous work. Firstly, that as the magnetic field is sheared, regions of increasingly high current density are created, evolving into thin current layers separating regions of potential open field \citep{1972ApJ...174..659B, 1986ApJ...309..383Y, 1991ApJ...380..655S}. In the presence of any resistivity, which must be present even if only at very high current densities, such strong current layers will eventually suffer reconnection and dynamic reconfiguration, involving the dissipation and ejection of magnetic energy \citep[e.g.][]{1991ApJ...382..677S}. In this scenario, the loss of equilibrium may or may not be associated with a ``critical'' twist amplitude (as the twist at which reconnection begins may or may not be independent of, or insensitive to, the form and strength of the triggering resistivity).

The second argument posits the existence of a critical twist angle, at which the magnetosphere experiences ideal magnetic non-equilibrium and evolves from a closed to a (partially) open configuration \citep[ML94;][]{2002ApJ...574.1011U}. This dynamic opening would be a purely ideal process; however, it results in the formation of a current sheet between oppositely directed open field lines, and so eventually must also be accompanied by reconnection-powered dynamics due to non-ideal physics in the sheet. 

It is of interest to determine whether the transition from quasi-static to dynamical evolution is consistent with there being a critical point, and if so to identify the critical twist amplitude $\psicrit$ at which the magnetosphere is reconfigured. In numerical simulations, it can be difficult to distinguish dynamic motion due to the (ideal or resistivity-driven) loss of equilibrium from the rapid progression through quasi-equilibrium states, because in a simulation twisting is applied on a finite timescale and at large $\psi$ the magnetosphere becomes very sensitive to further twisting. Furthermore, it can be difficult to distinguish the collapse of current layers to discontinuous current sheets due to ideal non-equilibrium from collapse caused by effective numerical resistivity when the layers' thickness approaches the grid scale. 

We choose to measure $\psirec$, the twist amplitude at which the fast reconnection phase begins, because this moment can be clearly defined: it is the first instant at which $E^2 > B^2$ anywhere in the domain (which necessitates the removal of electric field to mimic dissipation). We find that this inequality is satisfied only in the discontinuous current sheets that arise after the collapse of thicker current layers. A small amount of flux can reconnect before $\psirec$ due to effective resistivity in the current layer before it has fully collapsed; however this reconnection is slow and only involves a small fraction of the total reconnecting magnetic flux. If the shearing rate is slow then $\psirec$ should be approximately equal to $\psicrit$ (if a critical point exists); this is investigated and confirmed in \S~\ref{sec:shearrate}.

A critical twist angle may be determined by other means. For instance, ML94 defined $\psicrit$ as follows. After evolving to an ideal twisted configuration (which was still stable), they changed the equations of motion by introducing a resistive term $\eta \nabla^2\B$. If $\psi < \psicrit$, the twisted field lines slowly relaxed toward the initial potential state, while if $\psi > \psicrit$ the thick current layer at the equator collapsed, forming an X-point at which there was rapid reconnection of twisted field lines, and severing a large plasmoid of twisted flux which was ejected from the system.

\subsection{Equatorial shear}
\label{sec:eqshear}

Here we repeat the experiment of ML94, using the same surface shearing profile, Equation~(\ref{eq:MLprof}), but with relativistic force-free MHD. Rather than manually switching on magnetic diffusivity in the whole domain, we use spectral filters to consistently introduce resistivity in regions with very sharp field gradients (and hence high current density), as described in \S~\ref{sec:resist}. 

These simulations have grid size $N_r \times N_{\theta} = 384 \times 255$, and take place in a domain $1 \leq r \leq 60$, with the region $50 < r \leq 60$ comprising an absorbing zone. In order to keep the magnetosphere close to equilibrium, even at large shear, we evolve in three stages: from $\psi =$ 0 to 2.8, then from 2.8 to 3.6, and finally from 3.6 on, at shearing rates $\omega_0 = 1.4\times 10^{-3}$, $8\times 10^{-4}$, and $2.5\times 10^{-4}$ respectively. (We quote shears by the total azimuthal angular separation of the footpoints of the most twisted field line: $\psi = 2\int \omega_0(t) \,{\rm d}t$.) The magnetosphere is allowed to equilibrate between stages (although when twisting this slowly it anyway remains close to equilibrium). Increasing $\psi$ in three stages with decreasing twisting rate allows us to approach the critical point very gently.

\begin{figure}
\begin{center}
\includegraphics[width=85mm]{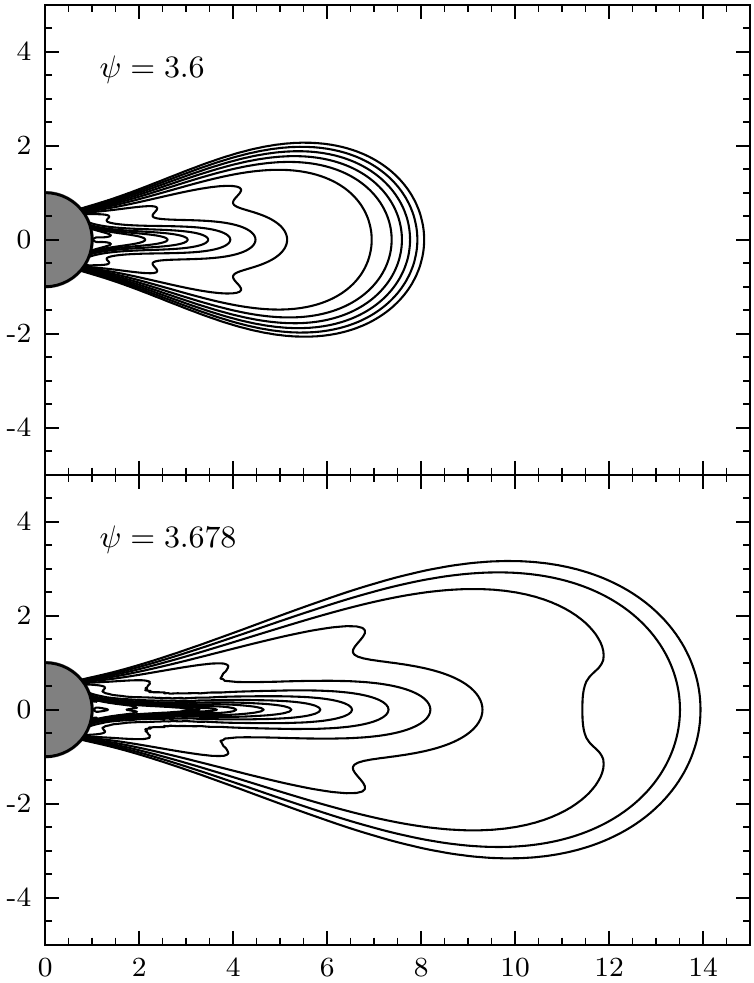}
\caption{Contours of toroidal current density, $J_\phi$, for the equilibrium state at $\psi =$ 3.6 and just before the onset of reconnection at $\psi =$ 3.678. The contours are equally spaced in $\ln\!\lp J_\phi\rp$ between $J_\phi = 10^{-4}$ and 0.25 $c\mu/\rstar^4$, with spacing $\Delta\ln\!\lp J_\phi\rp \approx 0.56$. The axes are labeled in units of $\rstar$. \label{fig:mljsheet}}
\end{center}
\end{figure}

The magnetosphere is still stable at $\psi =$ 3.6. As additional shear is applied, the field lines expand outwards and the current becomes more concentrated near the equator. The rates of expansion and concentration accelerate as $\psi$ increases, and eventually the current layer becomes sufficiently strong that non-negligible resistivity is introduced there by the spectral filters, and the layer's thickness suddenly decreases at $r \approx 3.4$, initiating reconnection. This occurs at a twist angle of $\psirec =$ 3.678. Contours of toroidal current density at $\psi =$ 3.6 and 3.678 are shown in Figure~\ref{fig:mljsheet}.

To test the sensitivity of this result to grid resolution, and hence to resistive (or reconnection) scale, we repeat the above procedure with a coarser $256 \times 155$ grid. Again the solution is indefinitely stationary at $\psi =$ 3.6, and reconnection begins at $\psirec =$ 3.654. Increasing the strength of the eighth-order filter (from $\alpha_{\rm SSV} =$ 0.025 to 0.05) and reducing the order of the high-order filter (from $2p =$ 36 to 26) each only reduces $\psirec$ by about 0.05\%. We conclude that the magnetosphere is extremely sensitive to shear above $\psi \approx$ 3.65, and that the point at which a discontinuous current sheet forms is insensitive to numerical parameters and the resistive length scale. This is consistent with the magnetosphere losing equilibrium at $\psicrit \sim$ 3.65.

These results are in approximate agreement with the findings of ML94; they estimated the critical shear for this profile to be roughly 4 radians, and found stable ideal configurations at this shear level. The difference may be due to their inclusion of gas pressure and the gravitational field of the sun, which resists the expansion of plasma, and hence field lines, away from the star.

Let us now consider the change in energy of the magnetosphere during the dynamical phase; the numbers given below are for the higher-resolution simulation (using the $384\times 255$ grid). At the onset of reconnection, the energy on the grid (in the domain $1 \leq r \leq 50$) is $W = 1.619\, W_0$, where $W_0$ is the energy of the dipole potential field in the same volume. The totally open field has $W \approx 1.662\, W_0$, and so the magnetosphere is, energetically speaking, about 94\% of the way from the dipole toward the limiting configuration (an underestimate, since more of the energy is stored at large radii beyond $r=50$ in the twisted state).

The dynamic evolution triggered at $\psi = \psirec$ involves the expulsion of magnetic energy from the system, in the form of a plasmoid of helical magnetic field disconnected by reconnection from the stellar surface, and the dissipation of magnetic energy in the current sheet.  The duration of the dissipative phase is $\Delta t_{\rm rec} \approx 60\,\rstar/c$.

During the dynamic phase the magnetosphere expels 53.68\% of the twist energy (of magnitude $0.619\, W_0$), dissipates 14.67\%, and retains 31.65\% in the form of static twisted flux. The result of doubling the strength of the eighth-order filter (to $\alpha_{\rm SSV} = 0.05$) and reducing the order of the high-order filter (from $2p_{\rm high} =$ 36 to 26) is given in Table~\ref{table:dissipfrac}. We can safely say that more energy is expelled than dissipated, but the precise contribution of each is dependent on numerical parameters; the amount of energy retained by the system is less sensitive to numerics. The dissipation fraction is predominantly controlled by the strength of the low-order filter, which removes energy from grid-scale features like current sheets. A precise measurement of the energy dissipated into heat by these reconnection events must await a resistivity prescription more closely modeling the relevant microphysics.

\begin{deluxetable}{l c c c}
\tablewidth{80mm}
\tablecaption{Fraction of the magnetic free energy retained, expelled, and dissipated during the dynamic phase, for three simulations with different spectral filtering parameters.\label{table:dissipfrac}}
\tablehead{  \colhead{$2p_{\rm high}$} &  \colhead{36} & \colhead{36} & \colhead{26} \\
\colhead{$\alpha_{\rm SSV}$} &  \colhead{0.025} & \colhead{0.05} & \colhead{0.025}   }
\startdata
retained & 31.65\% &  31.01\% & 31.65\% \\
expelled & 53.65\% & 51.84\% & 53.64\% \\
dissipated & 14.67\% & 17.15\% & 14.71\%
\enddata
\end{deluxetable}

\subsection{Polar cap shear}
\label{sec:pcshear}

In the preceding section we demonstrated, using one shearing profile, that at a certain twist amplitude the magnetosphere becomes extremely sensitive to further twisting, and that the critical point is largely insensitive to numerical parameters (such as grid resolution). In the following sections we investigate the critical point's dependence on the shearing rate, $\omega_0$, and on the surface shearing profile, $\omega(\theta)$. We use two grids, stretched in the radial direction using the exponential map, Equation~(\ref{eq:expmap}): the smaller grid has $N_r \times N_{\theta} = 640\times 255$ and an outer boundary at $\rout = 2155 \,\rstar$, the larger grid has a size of $1024 \times 375$ and $\rout = 2773 \,\rstar$. 

We will primarily study polar caps twisted at a constant rate until reconnection. Here, we illustrate the general behavior with reference to a single model, a cap extending to $\thpc = 0.15\pi$ twisted at $\omega_0 = 2.5\times 10^{-3}$. The calculation was performed using the larger grid. 

\begin{figure*}[tp]
\centering
\includegraphics[width=165mm]{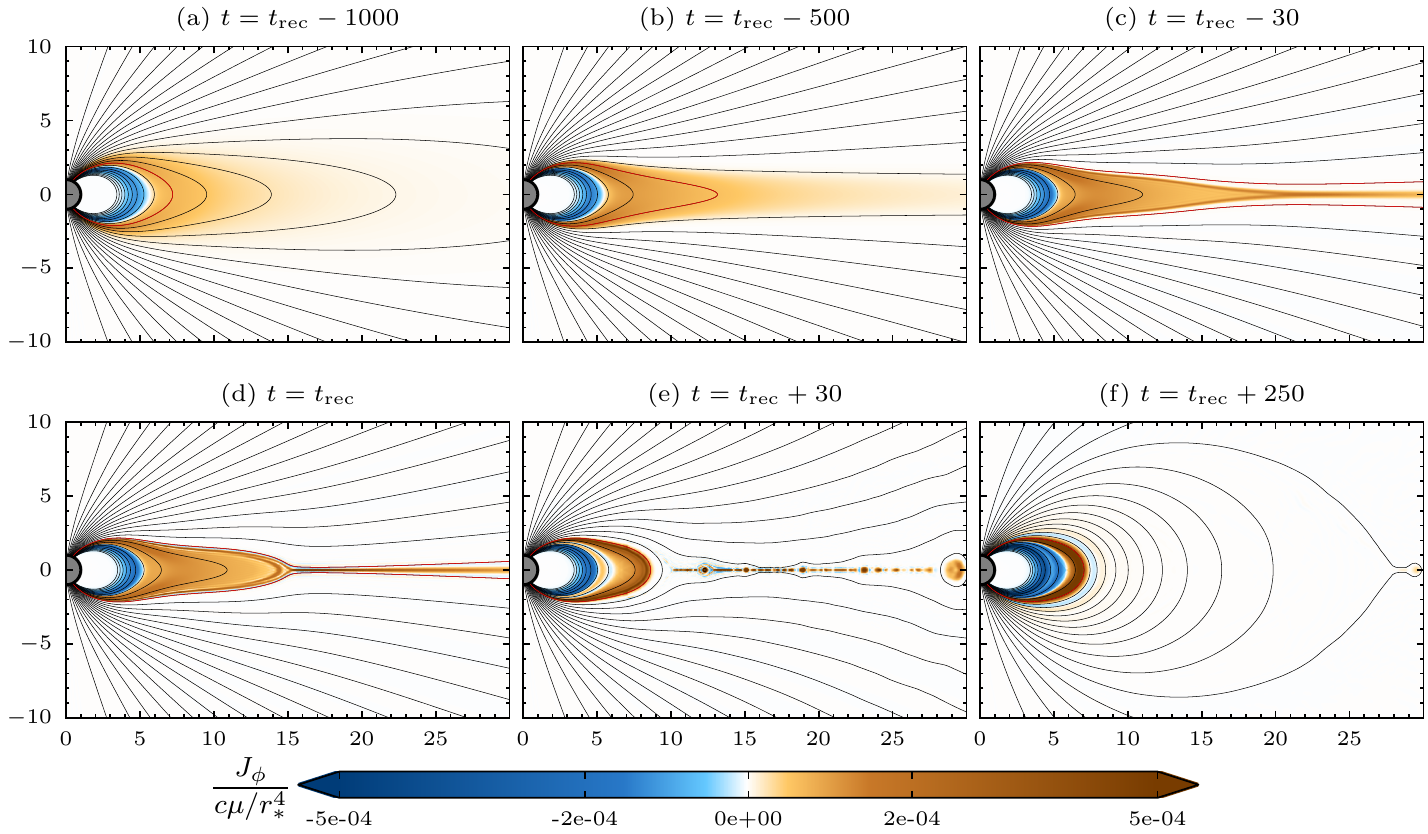}
\caption{Formation of the current sheet: states leading up to and following reconnection, for a polar cap of $\thpc = 0.15\pi$ and shearing rate $\omega_0 = 2.5\times10^{-3}\; c/\rstar$. The reconnection begins at $t_{\rm rec}$, panel (d). Color shows toroidal current density, lines are poloidal field lines projections, equally spaced in flux function in the range $u=$ 0.01--0.3 with spacing $\Delta u \approx 0.012$. One field line is highlighted in red; this field line first opens and then closes again when it reconnects. Time is in units of $\rstar/c$. \label{fig:polar-recon}}
\end{figure*}

The evolution of the magnetosphere through the reconnection event is shown in Figure~\ref{fig:polar-recon}. In Figure~\ref{fig:polar-recon}a, the magnetosphere is significantly inflated ($\psi \approx 3.1$) but a strong equatorial current layer has not yet formed. The magnetosphere is very sensitive to additional shearing at this point, and the current layer appears soon thereafter. By $\psi \approx 4.4$ (Figure~\ref{fig:polar-recon}b) almost all the flux which will eventually reconnect has been opened, yet the configuration is stabilized by the continual gradual expansion of closed field lines (see \S~\ref{sec:shearrate}).

The collapse of the current layer to a discontinuous sheet does not begin until $\psi \approx 5.5$ (Figure~\ref{fig:polar-recon}c). During the collapse, the Y-point-like cusp separating closed and open field lines retreats toward the star at speed $\sim c/4$, and the reconnection criterion is first satisfied at $\psi =\psirec = 5.59$, $t=t_{\rm rec}=2246$ (Figure~\ref{fig:polar-recon}d, in which the pinch point is clearly visible). The withdrawal of the closed flux further removes magnetic pressure support from the reconnection region, increasing the rates of collapse and reconnection.

Reconnection proceeds across the equatorial current sheet, producing plasmoids of a range of sizes, which are either ejected from the system or bounce backward and forward on closed field lines (Figure~\ref{fig:polar-recon}e). This process is in an approximately steady state for $\Delta t_{\rm rec} \approx 300\, \rstar/c$, with magnetic flux moving toward the current sheet at $v_\theta \sim 0.1\, c$, and the newly reconnected field lines rushing away from the reconnection point at $v_r \sim c$ in the equatorial plane. Unlike the outflow speed, which is determined by the large-scale magnetic tension of newly reconnected field lines, the flux inflow (or reconnection) speed may depend on the microphysical details of the reconnection process. The dimensionless reconnection rate in our simulations, $v_{\rm in}/v_{\rm out} = v_\theta/v_r \sim 0.1$, is similar to what has been found in particle-in-cell simulations of relativistic pair-plasma reconnection \cite[e.g.][]{2001ApJ...562L..63Z, 2011PhPl...18e2105L}.

Numerical dissipation eventually removes the trapped bouncing Alfv\'{e}n waves, and the magnetosphere relaxes to a steady state, retaining some twisted flux (Figure~\ref{fig:polar-recon}f). In this simulation, the configuration reaches this new equilibrium before any waves from the star interact with the outer boundary.

The stellar shearing has continued throughout this phase, although it is too slow to have much effect on the dynamical evolution described. After the dynamical phase, it begins to slowly re-twist the field lines which reconnection left purely poloidal. More importantly, the twist amplitude continues to grow on those field lines that did not reconnect and therefore did not lose any twist, which we will refer to as the ``twisted reservoir.'' Eventually the magnetic pressure in this strongly twisted region becomes so great that it explodes outward unstably, initiating a second reconnection event. The sudden explosive behavior is similar to the ``magnetic detonation'' described by \citet{1997PhR...283..185C}. 

\begin{figure}[tp]
\centering
\includegraphics[width=85mm]{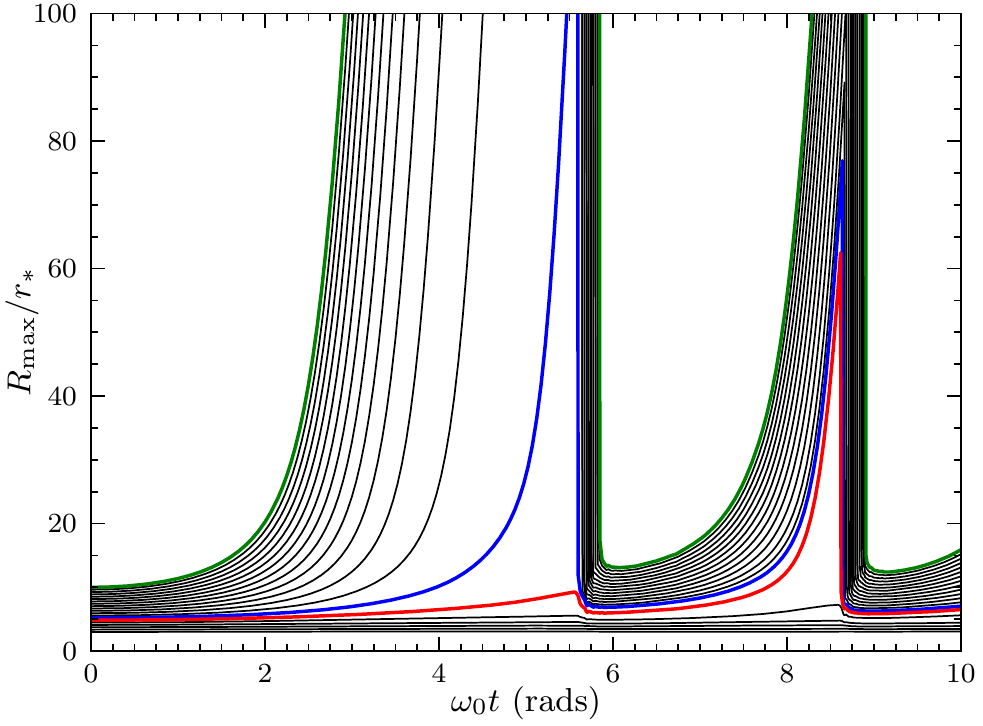}
\caption{Maximum field line heights versus applied shear, for the polar cap model twisted through two reconnection events. At $t=0$, the twenty field lines shown are equally spaced in $R_{\rm max}$, from $R_{\rm max}$ = 3 to 10 $\rstar$.  Three curves are highlighted with thicker colored lines. \label{fig:rmax-double}}
\end{figure}

The two rounds of overtwisting (inflation) and loss of twist (reconnection) are illustrated in Figure~\ref{fig:rmax-double}, which shows field line heights above the surface versus applied shear. For the second event, the twist amplitude in the flux which had previously reconnected is much lower than the total twist accumulated in the twisted reservoir, and so the evolution is more like that of a twisted ring solution than a polar cap model. The weakly twisted overlying flux acts as a kind of nozzle, keeping the expanding flux near the equator, which causes it to reconnect earlier (see \S~\ref{sec:shearprof}). For example, the field line indicated by the blue line in Figure~\ref{fig:rmax-double} reconnects when its apex is at $R_{\rm max} \approx 140$ in the first event, but it only reaches $R_{\rm max} \approx 75$ in the second event. The red line in Figure~\ref{fig:rmax-double} represents a field line which expands but remains closed during the first event (the drop in $R_{\rm max}$ is due to the rapid retraction of the closed bundle) and experiences reconnection in the second event, during which more field lines expand and reconnect because the expansion is driven by deeper flux.

In both events shown in Figure~\ref{fig:rmax-double}, reconnection of the field lines shown takes $\Delta t \approx 100\, \rstar/c$. Reconnection continues beyond this time on field lines having $R_{\rm max}(t \! =\! 0) > 10$ (not shown); however, most of the energy will be released by the lowest-lying reconnecting field lines. The entire reconnection phase takes about $\Delta t_{\rm rec} \approx$ 200--300 $\rstar/c$. In general, we find that the reconnection timescale is approximately given by
\beq
\Delta t_{\rm rec} \sim \frac{2 R_{\rm rec}}{v_{\rm rec}}
\eeq
regardless of the shearing model, where $R_{\rm rec}$ is the initial inner radius of the reconnecting current sheet and $v_{\rm rec} \sim 0.1 c$. 

At the onset of the first reconnection event the magnetosphere has free energy $W_{\rm tw} = 0.0069\, W_0$ in the volume $1 < r < 500$. During the dynamic phase 9.9\% of the free energy is expelled from the system, 43.8\% is dissipated, and 46.3\% is retained in the new quasi-equilibrium state. This dissipation fraction is approximately three times as large as was found in the fiducial equatorial shearing model in Section~\ref{sec:eqshear}.

\subsection{Dependence of $\psirec$ on shearing rate}
\label{sec:shearrate}

\begin{figure}[tp]
\centering
\includegraphics[width=85mm]{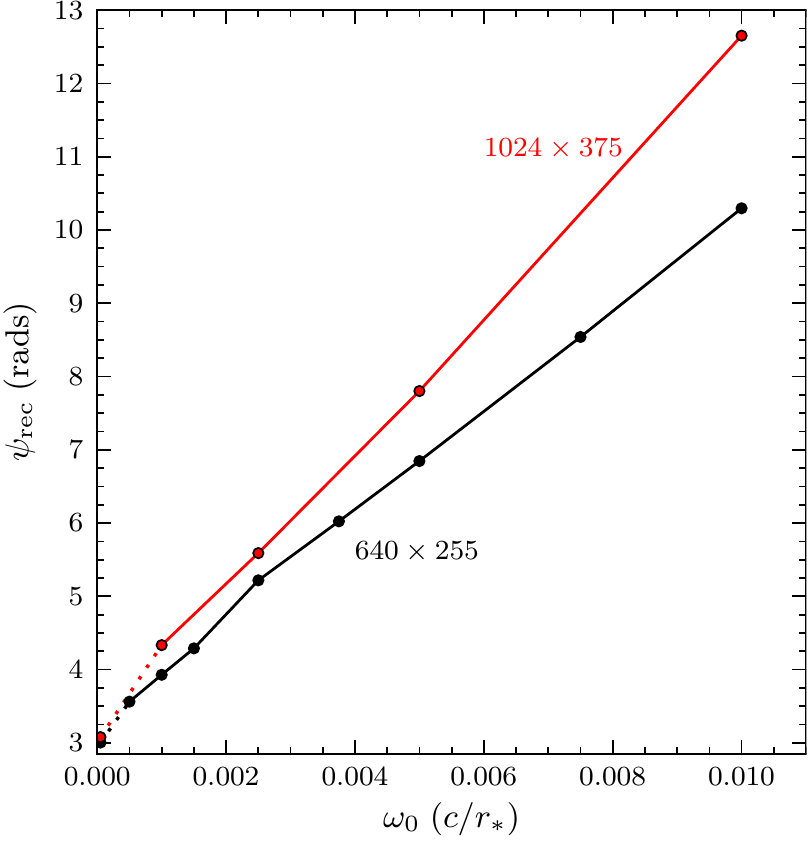}
\caption{Twist angle at onset of reconnection versus shearing rate, for a polar cap with $\thpc = 0.15\pi$. The points at smallest $\omega_0$, connected to the others with dotted lines, were found by shearing in stages with progressively lower $\omega_0$. Results marked in black were obtained using the smaller grid ($640\times 255$), those shown in red were obtained with the larger grid ($1024\times 375$).\label{fig:psi-omega}}
\end{figure}

We now turn to the effect of the shearing rate, which we study with the shearing profile employed in the preceding section, a polar cap extending from the northern axis to $\thpc = 0.15 \pi$. In these simulations the shearing rate is smoothly increased from zero to $\omega_0$ and then held constant. In Figure~\ref{fig:psi-omega}, the twist angle at which reconnection begins, $\psirec$, is plotted against $\omega_0$ for shearing rates from $\omega_0 = 5\times 10^{-4}$ to $10^{-2}$.

The twist angle at reconnection is a strong, approximately linear, function of the shearing rate,
\beq
\psirec \approx \psirec^0 + {\rm const}\cdot \omega_0.
\label{eq:psi-omega}
\eeq
The magnetospheric instability is delayed to large $\psirec$ at high twisting rates $\omega_0$. This ``dynamical stabilization'' appears to be the result of a competition between the formation of a thin current layer following the loss of equilibrium at $\psicrit$ and field inflation driven by twisting (see Figure~\ref{fig:polar-recon}). The thin current layer appears first at a pinch point, just outside those closed field lines which do not inflate rapidly and will not undergo reconnection. Near this point the expansion of the opening field lines forms a region of lower magnetic pressure. The formation of the current layer is slow at first and becomes faster as it proceeds. In its early stages, the narrowing of the high-current region may be counteracted by continual twisting, which causes the last closed field lines to expand, pushing them through the pinch point and preventing the overlying open lines from moving toward the current layer. The rate at which closed flux is pushed through the pinch point decreases as more flux becomes open, and the thickness of the current layer decreases slowly. Eventually the current layer's thickness approaches the grid scale and it becomes resistive. The resistivity causes the layer to collapse quickly to a discontinuous current sheet, initiating reconnection. The point at which this occurs, for a given shearing profile, will depend on the shearing rate and the resistive length scale set by the numerical grid.

Equation~(\ref{eq:psi-omega}) explains the deviation of the total magnetic energy from the virial theorem energy at large twist amplitude in Figure~\ref{fig:energy_equil}b---the magnetosphere is in the dynamically stabilized state for $\psi \gtrsim 3$. This is confirmed by further simulations, in which we gently implant a certain twist amplitude, the shearing rate being smoothly decreased such that it becomes zero at $t_{\rm tw}$ when the total shear reaches $\psi_{\rm tw}$. We find that the magnetosphere is stable after being twisted to $\psi_{\rm tw} = 3.1$, while current sheet formation and reconnection occur if $\psi_{\rm tw} = 3.25$.

In Figure~\ref{fig:psi-omega}, the curves for the two numerical grids converge as $\omega_0$ is decreased---in the limit of quasi-static twisting, $\psirec = \psicrit$ is independent of grid resolution and hence resistive length scale. The dynamical stabilization disappears as $\omega_0 \rightarrow 0$. The slope of each $\psirec(\omega_0)$ curve is determined by the resistive length scale of the simulations (here roughly the grid scale). As the resistive length scale is decreased, the dynamical stabilization is effective to larger $\psi$ because higher field gradients must be created before the current layer will resistively collapse, and so the slope $\Delta \psirec/\Delta\omega_0$ increases with grid resolution.

A finite shearing rate complicates the picture of a magnetosphere evolving quasi-statically toward a well-defined unstable critical point. Any twisting rate, even one which keeps the magnetosphere almost perfectly in equilibrium up to large $\psi$, as in Figure~\ref{fig:energy_equil}, eventually becomes dynamically important, delaying the onset of reconnection. Each model represented in Figure~\ref{fig:psi-omega} undergoes reconnection (on a timescale shorter than the twisting timescale) if the surface shearing is smoothly halted above a twist of $\psi \sim 3$. Beyond this angle they can only temporarily be in dynamically stabilized states---there are no corresponding true equilibrium configurations to which they may gently relax. 

The stabilizing effect of twisting can be thought of as being analogous to the effect of pedaling when riding a bicycle---one becomes more stable against spontaneous toppling when pedaling more quickly. Of course, faster pedaling may lead one more quickly to a catastrophic end---such as a collision---whose intensity is amplified by the greater speed. Similarly, a higher twisting rate will drive a magnetosphere to large-scale reconnection in a shorter time (though at a larger twist amplitude), and increase the total amount of magnetic free energy stored, and then expelled and dissipated in the dynamic phase.  For example, in the simulations described above the free energy of the magnetosphere at the onset of reconnection is about $2.2$ times larger for $\omega_0 = 10^{-2}$ than for $\omega_0 = 10^{-3}$ ($1.39\times 10^{-2}\, W_0$ versus $6.27\times 10^{-3}\, W_0$, both found using the larger grid). 

As $\omega_0 \rightarrow 0$,  the measured $\psirec$ should go to the critical angle $\psicrit$.  To study the behavior in the very slow twisting limit we evolved the magnetosphere in a series of stages, with decreasing $\omega_0$, in a manner similar to \S~\ref{sec:eqshear}; the configuration is allowed time to equilibrate between stages. In the final stage, we evolve from a stable equilibrium state through the point of reconnection, with a shearing rate of $\omega_0 = 5\times 10^{-5}$. Using the smaller grid, we bring the configuration from a stable state at $\psi=2.75$ to reconnection at $\psirec = 3.005$; with the larger grid, we find a stable equilibrium at $\psi = 2.95$ and measure reconnection at $\psirec = 3.079$. These measurements (shown in Figure~\ref{fig:psi-omega}, connected to the constant-twisting values by a dotted line), agree to a fractional discrepancy of about two percent, confirming that at low twisting rates $\psirec$ is insensitive to numerical resolution and the details of the spectral filtering.

These measurements may slightly underestimate $\psicrit$. Because of the slow twisting rates and time allowed for equilibration, the outer boundary is not out of causal contact with the star at all times, and eventually outgoing waves on a fraction of the very inflated field lines are removed by the absorbing layer. This artificially reduces the magnetic pressure near the boundary, drawing out the inflated flux, and generally causing the configuration to be further along the path to current sheet formation than it would be naturally at a given $\psi$. We have confirmed this effect using domains with $\rout \sim 100$--250, finding that decreasing $\rout$ leads to reconnection at smaller $\psirec$. This effect may explain why the very-slow-shearing values lie below the lines formed by the constant-twisting measurements, and may be responsible for the kink in the black curve (for the smaller grid) at $\omega_0 = 2.5\times 10^{-3}$ in Figure~\ref{fig:psi-omega}---below this shearing rate, the outer boundary is no longer out of contact for the whole simulation\footnote{In the following section we find stable solutions for the $\thpc=0.15\pi$ polar cap problem at $\psi = 3.1$ when the entire simulation takes places before any waves reach the boundary, while, as described above, reconnection occurs at $\psi = 3.079$ when the simulation is long and outgoing waves are removed at the boundary (these simulations were performed using the same computational grid and spectral filters).}.

\subsection{Current sheet formation}
\label{sec:collapse}

When the shear applied to a magnetosphere is slowly brought to the critical twist amplitude the field lines expand outward to large heights above the stellar surface. As the field lines expand the current layer separating expanding flux of opposite directions (or sign of $B_r$) becomes thinner, tending toward a discontinuous current sheet. This field line expansion is an ideal MHD process.

The critical behavior can also be seen if a magnetosphere is twisted past the critical angle at a finite shearing rate, and the shearing rate is then slowly reduced to zero. Since no steady state is available at $\psi > \psirec$ the magnetosphere must become unstable. We use the toroidal current density $J_\phi$ to measure the strength of the current layer at the equator; a discontinuous current sheet has infinite $J_\phi$. Using the larger $1024\times 375$ grid, a polar cap with $\thpc = 0.15\pi$ is brought to a maximum twisting rate of $\omega_0 = 2.5\times 10^{-3}$, which is held constant for approximately $\Delta t = 10^3\, \rstar/c$ and then reduced slowly to zero by $t_{\rm tw}$, such that a total twist amplitude of $\psi_{\rm tw} = 3.25$ is implanted by $t_{\rm tw} = 1690\, \rstar/c$. For $t < t_{\rm tw}$, $J_\phi$ grows approximately linearly with time. Following $t_{\rm tw}$, the current density continues to increase, albeit more slowly, even though no further shearing is applied. About $100$--$200\,\rstar/c$ after $t_{\rm tw}$ , the current layer's thickness approaches the grid scale, resistivity becomes dynamically important, and the thickness of the current layer dives increasingly quickly to zero, giving ${\rm d}^2 J_\phi/{\rm d} t^2 > 0$.

The collapse and reconnection process is consistent with the tearing instability \citep{1963PhFl....6..459F, 2003MNRAS.346..540L, 2007MNRAS.374..415K}, which begins with slow or linear evolution as the instability threshold is crossed, eventually enters a rapid evolution phase, and produces a large number of plasmoids of a range of sizes.

\begin{figure}[tp]
\centering
\includegraphics[width=85mm]{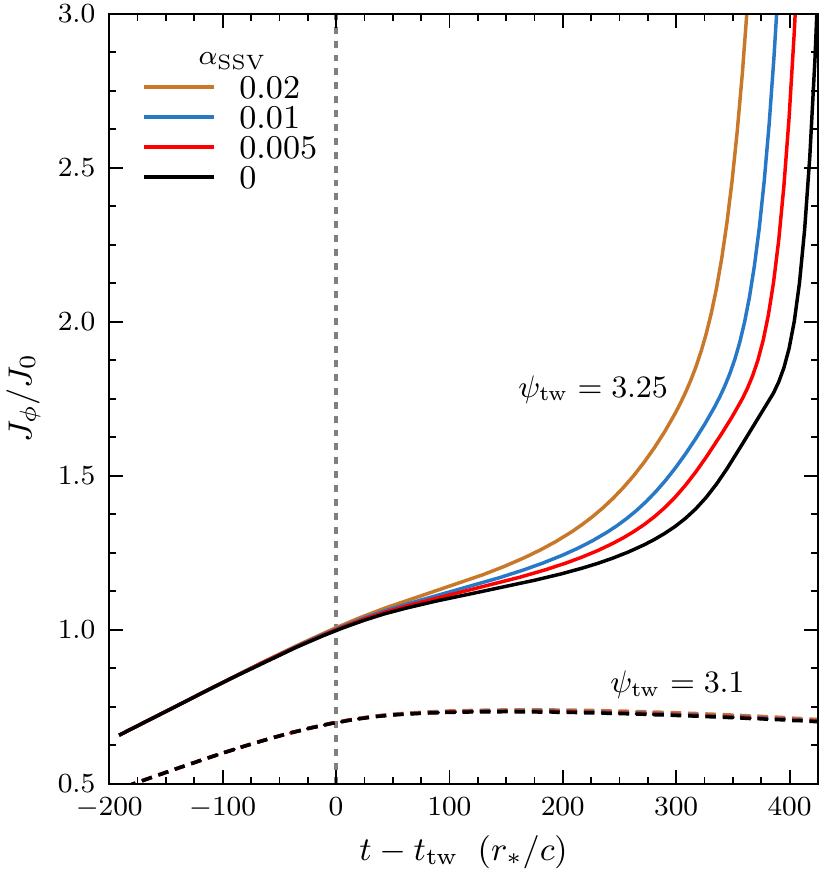}
\caption[Collapse to a current sheet.]{Collapse to a current sheet, showing $J_\phi$ at $(r,\theta) = (40, \pi/2)$, in units of its value at $t = t_{\rm tw}$ ($J_0 = 4.09\times 10^{-5}$ $c\mu/\rstar^4$), for different values of the filtering strength $\alpha_{\rm SSV}$. With a final implanted twist amplitude of $\psi_{\rm tw}=3.25$ (solid curves) the evolution becomes sensitive to the effective resistivity level after the shearing rate reaches zero at $t_{\rm tw}$. The dashed curves represent simulations having a maximum twist angle of $\psi_{\rm tw} = 3.1$; here the evolution remains insensitive to the filtering strength. \label{fig:collapse}}
\end{figure}

The ideal and resistive phases are illustrated in Figure~\ref{fig:collapse}, which shows $J_\phi$ on the equator at $r=40$, approximately the radius at which the X-point first begins to form, for four values of the filtering strength $\alpha_{\rm SSV}$ (a higher value implies more resistivity, and when $\alpha_{\rm SSV}=0$ the numerical resistivity is due to the high-order filter; see Section~\ref{sec:resist}). For $t < t_{\rm tw}$ the evolution is independent of the resistivity level. After this time the models diverge, the current layers in those with higher resistivity collapsing earlier toward a discontinuous current sheet. It is possible that there is a slow ideal collapse phase, and that in the perfect absence of resistivity the magnetosphere would still gradually evolve toward a state with open field lines and an infinitely thin tangential magnetic field discontinuity. However, at the critical point the magnetosphere becomes extremely sensitive to resistivity, and we were unable to isolate a filtering- or resistivity-independent collapse phase. In all our simulations---both using the equatorial shearing and the polar cap twisting models---the identifiable collapse behavior occurs when the current layer thickness is near the grid scale; increasing grid resolution merely slightly increases the twist angle at which resistivity takes over and initiates collapse.

\begin{figure}[tp]
\centering
\includegraphics[width=85mm]{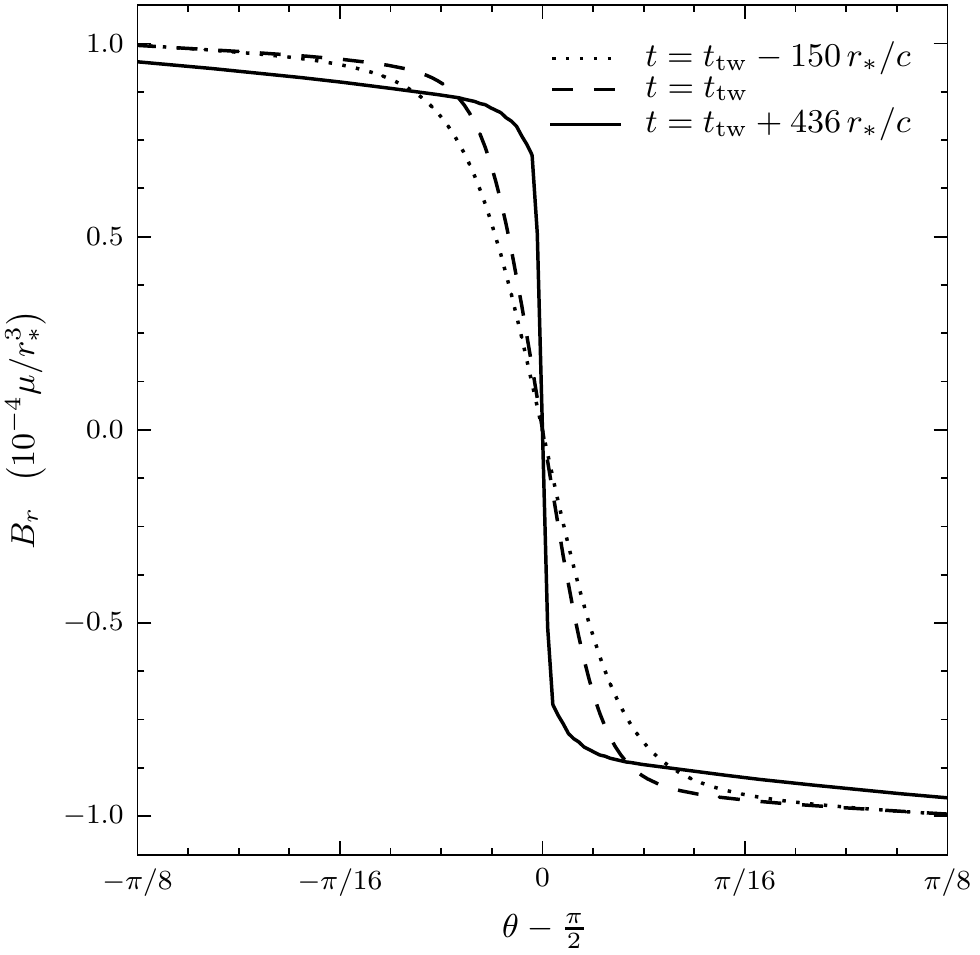}
\caption[$B_r$ jump during current sheet formation]{Jump in $B_r$ across the equator during the formation of the current sheet, at $r = 40\, \rstar$, from the $\alpha_{\rm SSV} = 0.005$ simulation.
The dashed curve shows the current layer when the shearing rate reaches zero, $t = t_{\rm tw}$. The jump becomes unresolved by the grid at $t=t_{\rm tw}+436\, \rstar/c$ (solid curve). \label{fig:brjump}}
\end{figure}

The $\alpha_{\rm SSV} = 0.005$ model (red curve in Figure~\ref{fig:collapse}) begins to collapse rapidly at $t \approx t_{\rm tw} + 200$; in this simulation the jump in $B_r$ across the equator becomes unresolved by the grid at $t\approx t_{\rm tw} + 436$ (see Figure~\ref{fig:brjump}). The X-point moves inward as the current sheet becomes stronger (see Figure~\ref{fig:polar-recon}), and the reconnection criterion $E^2>B^2$ is first satisfied at $t_{\rm rec} = t_{\rm tw} + 500.1$ and $r = 24.3$. The collapse and subsequent reconnection occur before any waves from the star have reached the outer boundary.

Figure~\ref{fig:collapse} also illustrates how sensitive the magnetosphere becomes to shearing at large twist amplitudes. At $t = t_{\rm tw} - 100$ the shearing rate is only $1.4\times 10^{-4}$ (and is slowly decreasing), yet the current density at the equator grows strongly. Once the shearing is withdrawn completely the current density growth slows, but under the influence of resistivity the magnetosphere continues to evolve toward a partially open state (i.e. a state with some closed and some open field lines, and a sharp current sheet separating open field lines of opposite polarity). At the filtering levels shown ($\alpha_{\rm SSV}=0$--$0.02$), our code applies very little dissipation on the scales at which the resistive collapse begins. The configuration is very sensitive to resistivity, and even the small amount introduced by the spectral filters is enough to initiate the resistive instability. As the layer becomes thinner it becomes more resistive, and so the collapse accelerates unstably.

If $\psi_{\rm tw}$ is below the critical amplitude, the magnetosphere instead gently relaxes toward a stable equilibrium. This is shown in Figure~\ref{fig:collapse} for $\psi_{\rm tw} = 3.1$, with colored dashed lines representing the same four resistivity levels. These lines nearly lie on top of one another, implying that the relaxation is an ideal process. (Eventually, when $t \gtrsim t_{\rm tw} + 600$, the equilibrium begins to resistively diffuse, and the different $J_\phi(t)$ curves begin slowly to diverge depending on the filtering level.)

\subsection{Dependence on shearing profile}
\label{sec:shearprof}

\begin{figure}
\centering
\includegraphics[width=85mm]{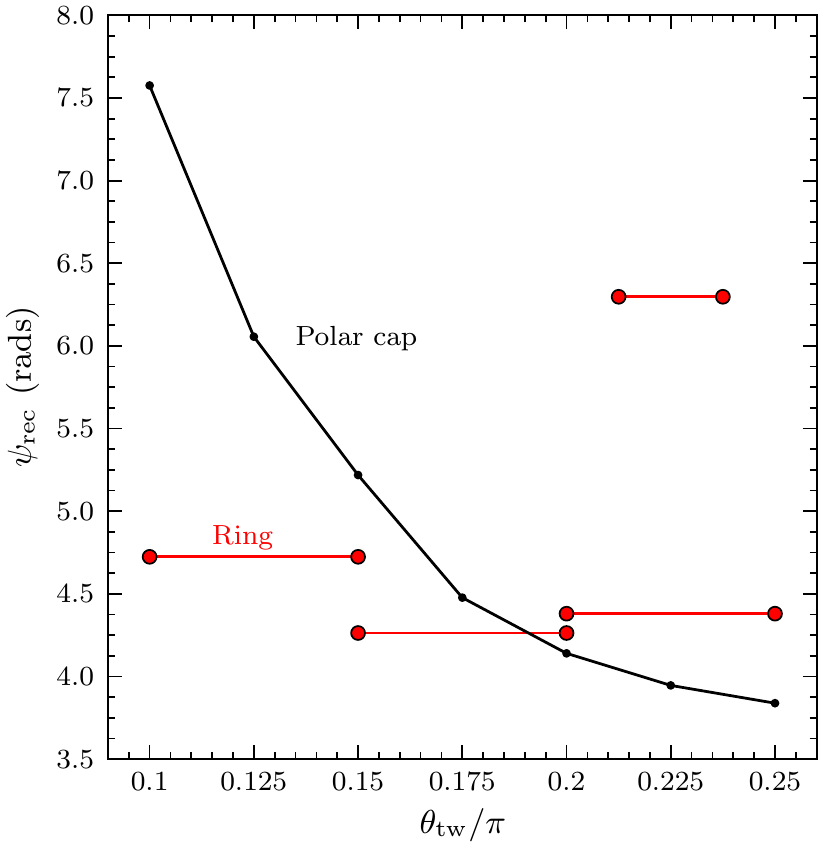}
\caption{ Twist angle at onset of reconnection for various shearing profiles, for shearing rate $\omega_0 = 2.5\times 10^{-3}\; c/\rstar$. The horizontal axis denotes $\thpc$ for the polar cap models (black curve); for the ring models, the red horizontal lines begin and end at $\theta_1$ and $\theta_2$ respectively. \label{fig:psi-profile}}
\end{figure}

To investigate the effect of changing the profile of the applied surface shear, we performed several simulations (using the smaller grid), at the same shearing rate $\omega_0 = 2.5\times 10^{-3}$, for polar caps of different sizes and rings of different central latitudes and widths. Figure~\ref{fig:psi-profile} plots $\psirec$ for these models (for reference, the black point at $\thpc = 0.15\pi$ refers to the same simulation as the black point at $\omega_0 = 2.5\times10^{-3}$ in Figure~\ref{fig:psi-omega}).

The polar cap models show a strong dependence of $\psirec$ on $\thpc$. Very large twist angles are required to induce reconnection for small polar caps. Loosely speaking, the energy required to open a flux bundle centered on $f$, $W_{f}$, increases more rapidly with flux function near the equator than near the pole: ${\rm d^2} W_{f}/{\rm d^2} f > 0$. Therefore continual twisting can dynamically stabilize a flux bundle at smaller $f$ more effectively against collapse and reconnection. However, this is a dynamic effect (described in Section~\ref{sec:shearrate}). When the procedure of evolving increasingly slowly in stages is used, a polar cap model with $\thpc = 0.1\pi$ undergoes reconnection at $\psirec = 3.086$, close to the value given in the preceding section ($\psirec = 3.005$, for a model with $\thpc = 0.15\pi$; both simulations used the $650\times 255$ grid). In the quasi-static limit, the critical twist angle for a sheared polar cap is roughly independent of its size. This is not surprising given the self-similarity of the dipole field, since changing $\thpc$ is equivalent to changing the radius of the star.

The ring models are more complicated. They have untwisted field blanketing the twisted flux; this untwisted field has no $B_{\phi}^2$ pressure, and would not naturally undergo poloidal expansion. The untwisted flux acts to confine the twisted field underneath, which must push against its magnetic tension in order to inflate. Consider a ring extending from $\theta_1$ to $\theta_2$, where $\theta_1$ is the colatitude of the footpoint nearer the pole. Equilibrium ring solutions, at a given $\psi$, have decreasing poloidal expansion with increasing $\theta_1$; in particular, they have inflated less than a polar cap model with $\thpc = \theta_2$ (a polar cap can be thought of as a ring with $\theta_1 = 0$). Magnetic pressure builds up in the weakly expanding twisted flux bundle; when the configuration loses equilibrium, this magnetic overpressure region can inflate explosively at nearly the speed of light. This explosive mode of field opening will be discussed further in \S~\ref{sec:rotate}.

There are three principal effects determining $\psirec$ for a twisted ring. 
\begin{enumerate}
\item The untwisted flux acts like a nozzle, compressing the expanding flux around the equator. This prevents the dynamic stabilizing mechanism from operating as effectively, because closed field lines need to push overlying open lines out of the way in order to open themselves. Ring solutions therefore tend to experience reconnection at angles closer to their quasi-static critical angle than polar cap solutions; this is why, in Figure~\ref{fig:psi-profile}, the ring extending from $0.1\pi$ to $0.15\pi$ reconnects earlier than the polar cap with $\thpc = 0.15\pi$. 

\item Decreasing the size of the ring increases the twist required for instability and reconnection, because narrower rings inject less energy into the magnetosphere; this is illustrated by the two rings centered at $\theta_{\rm ctr} = 0.225\pi$. 

\item $\psirec$ increases as a ring is moved closer to the equator, because larger twist angles are needed to inflate high-tension deeply buried field lines, and to push away increasingly ``heavy'' enveloping untwisted field; this explains why the ring from $0.2\pi$ to $0.25\pi$ reconnects at a larger twist angle than the ring extending from $0.15\pi$ to $0.2\pi$ (even though it contains 11\% more twisted flux). 

\end{enumerate}

\section{Rotating stars}
\label{sec:rotate}

In this section, we study the evolution of the differentially twisted magnetosphere of a star that is already in solid body rotation. This problem was first examined numerically by \citet{2012arXiv1201.3635P}, where particular consideration was given to the effect of twisting on the star's spindown magnetic braking rate. Here we present a more thorough investigation, placing the twisting rotating solutions in the context of the non-rotating configurations described above.

Stellar rotation adds a new characteristic scale to the problem. Spatially, this scale is the light cylinder radius, $\rlc$, the distance from the rotation axis at which corotation with the stellar surface implies motion at the speed of light: $\rlc \equiv c/\Omega$. No field lines can close outside the light cylinder in an ideal magnetosphere; rotation therefore puts an upper limit on the possible size of a stable bundle of closed flux. Beyond the light cylinder, field lines are open to infinity. The new characteristic scale can also be viewed as the flux function of the last open field line, $\urot = \sin^2\throt$, where $\throt$ is the colatitudinal extent of the polar cap of open flux, $\throt \approx \lp\rstar/\rlc\rp^{1/2}$.

Twisting of open field lines simply results in the twist energy flowing away from the system. More interesting effects appear if surface shearing twists field lines at larger $u$, in the closed zone of the magnetosphere. Each twisting, rotating model can be characterized by the dimensionless number $a$, which relates the amount of twisted flux to the rotationally opened flux, $\urot$. For a polar cap model, it is simply the ratio of the magnetic fluxes emerging through the polar caps $\theta < \thpc$ and $\theta < \throt$,
\beq
a = \frac{\upc}{\urot} = \frac{\sin^2\thpc}{\sin^2\throt};
\eeq
all models considered below have $a>1$.  Similarly, a ring model is described by two numbers, $a_1$ and $a_2$, which label the two flux surfaces that bound the sheared flux bundle in units of $\urot$; the model will be referred to as ``$a_1$--$a_2$.'' A ring extending from $\theta_1$ to $\theta_2$ is labeled by $ a_1 = \sin^2 \theta_1/\sin^2\throt$ and $a_2 = \sin^2 \theta_2/\sin^2\throt$; the fractional magnetic flux through the twisted ring is $(a_2-a_1) \urot$.  

Two numerical grids are employed in the simulations described in this section, depending on the light cylinder radius. The first grid has a resolution of $N_r \times N_\theta = 384 \times 255$ and is used when $\rlc = 20\, \rstar$, the second has $768\times 507$ grid points and is used when the rotation rate is half as fast, $\rlc = 40\, \rstar$. Increasing the light cylinder radius allows one to twist a larger multiple of the rotationally-opened flux, which shrinks as $\rlc$ increases. In particular, all simulations with $a$ or $a_2$ greater than 7 are performed with the $\rlc = 40\, \rstar$ grid.

At each instant of time, there are two further numbers which quantify the changing configuration. The first is the rate at which energy is transferred from the star to the magnetosphere; this is the Poynting flux integrated over the stellar surface,
\beq
L =  \frac{\rstar^2}{2} \int_0^{\pi} \lp \E\times\B\rp_r \sin\theta\, {\rm d}\theta.
\label{eq:L}
\eeq
The second number is the torque on the star applied by the magnetosphere (the magnetic braking rate), which is the integrated angular momentum flux,
\beq
T = -\frac{\rstar^3}{2} \int_0^{\pi} \lp E_r E_\phi + B_r B_\phi \rp \sin^2\!\theta\, {\rm d}\theta.
\label{eq:T}
\eeq
In a steady state, the torque is applied solely by open magnetic flux\footnote{During twisting, energy and angular momentum are transferred from the star to the closed zone in the magnetosphere, where they are stored.}. Shearing increases the torque on the star because it pushes previously closed field lines through the light cylinder. This is a strong effect at large $\psi$, which we will study with several models in the sections which follow. We label the Poynting luminosity of, and magnetospheric torque on, the untwisted rotating star by $L_0$ and $T_0$ respectively.

\subsection{Quasi-steady twisted rotating states}

\begin{figure*}[tp]
\centering
\includegraphics[width=165mm]{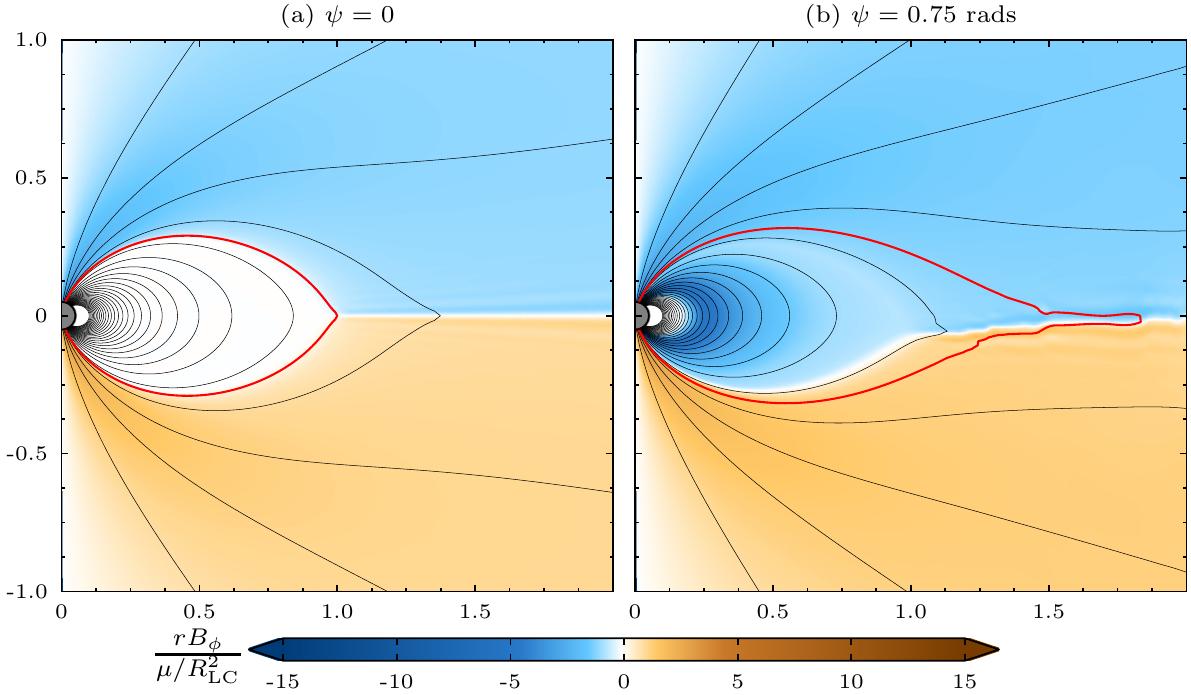}
\caption{ (a) Initial solution: untwisted rotating star; (b) quasi-steady twisted magnetosphere. Color shows toroidal magnetic field times the radial coordinate, in units where this quantity is initially $\sim 1$ at the light cylinder; 
poloidal field lines, 30 equally spaced in flux function between $u =$ 0.01 and 0.5, are shown in black; an additional field line, that initially closes at the light cylinder, is marked in red.  Axes are labeled in units of $\rlc = 20\,\rstar$. \label{fig:rot-early}}
\end{figure*}

We first consider the combined effects of twisting and solid-body rotation on the sequence of twisted force-free equilibria. Before the shearing is begun, we construct the equilibrium untwisted rotating configuration. A star is brought from rest to a constant rotation rate of $\Omega = 0.05$ (giving $\rlc = 20\,\rstar$) over several stellar light-crossing times, and the solution is allowed to relax to a steady state (Figure~\ref{fig:rot-early}a) as described by \citet{2012MNRAS.423.1416P}. This rotating configuration has fractional open flux of $\urot \approx 0.071$. Surface shearing is then applied to this equilibrium state on top of the stellar rotation. Starting at $t = 0$, we shear a polar cap, the magnetic flux through which is three times the flux opened by rotation (i.e. $a=3$). This cap is slowly brought from corotation to a maximum shear angular velocity of $\omega_0 = \Omega/50$.  The twisted polar cap extends down to $\thpc = \sin^{-1}\!\lp3\urot\rp^{1/2} \approx 0.153\pi$, which is comparable to the polar cap model studied in \S~\ref{sec:shearrate}. 

As the surface is sheared, Alfv\'{e}n waves are injected onto the field lines rooted in the twisted cap, causing them to bend in the azimuthal direction. Those waves which travel along open field lines flow out of the system at the speed of light, and the toroidal field on these lines does not increase significantly beyond its rotationally induced value. On the other hand, waves on closed field lines are trapped, and so the total twist increases on the twisted closed flux, storing energy in the magnetosphere. As in the non-rotating twisted solutions (see Figure~\ref{fig:equil_solns_2}), the additional magnetic pressure from the toroidal field causes these twisted closed lines to expand outward, pushing some previously closed field lines through the light cylinder (Figure~\ref{fig:rot-early}b).  

When a field line expands through the light cylinder, it initially remains closed in the equatorial current sheet due to finite resistivity there; it opens completely to infinity with only a small amount of further poloidal expansion. The newly opened field line has $B_\phi$ of opposite signs at its two footpoints, and so makes a net contribution to $T$. Thus, as more field lines are pushed outside $\rlc$ the spindown torque increases. Field lines undergo erratic reconnection as they are pushed through the resistive current sheet, sometimes creating discrete plasmoids which outflow through the sheet. The solution\footnote{The bulging of the closed lines above, but not below, the equator in Figure~\ref{fig:rot-early}b occurs because only the field lines' northern footpoints are sheared. The twisting causes $B_\phi$ to increase by a small amount on northern open field lines---they expand slightly and pull the closed flux below along with them. This effect becomes less pronounced as $\omega_0/\Omega$ is decreased; it is not evident in Figure~\ref{fig:rot-pseudo-crit}, because by the times shown the shearing has been halted and the extra twist energy on open lines has propagated out of the domain.} at a given $\psi$, such as in Figure~\ref{fig:rot-early}b, is quasi-steady, in that most of the flux (and integrated quantities like $L$ and $T$) are stable, but a small fraction of the field lines (those closing in the sheet) continue to fluctuate on short timescales.

\subsection{Overtwisting of rotating magnetospheres}
\label{sec:rot-crit}

In this section we describe three simulations, in which twists of three different amplitudes, $\psi_{\rm tw} =$ 1, 1.5, and 1.75, are carefully implanted in a rotating magnetosphere. We begin with the same steady-state rotating solution as in the previous section. At $t = 0$, a polar cap with $a=3$ is brought from corotation to a shear angular velocity of $\omega_0 = \Omega/50$. The shearing rate is held constant for some time (on the order of ten rotational periods), and then smoothly reduced to zero over a similar time scale, giving a final twist amplitude of $\psi_{\rm tw} = \int_0^{t_{\rm tw}}\omega_0(t)\,{\rm d}t$, where $t_{\rm tw}$ is the time at which the shearing rate returns to zero. The rotating configuration is then evolved with no further shearing.

\begin{figure}
\centering
\includegraphics[width=85mm]{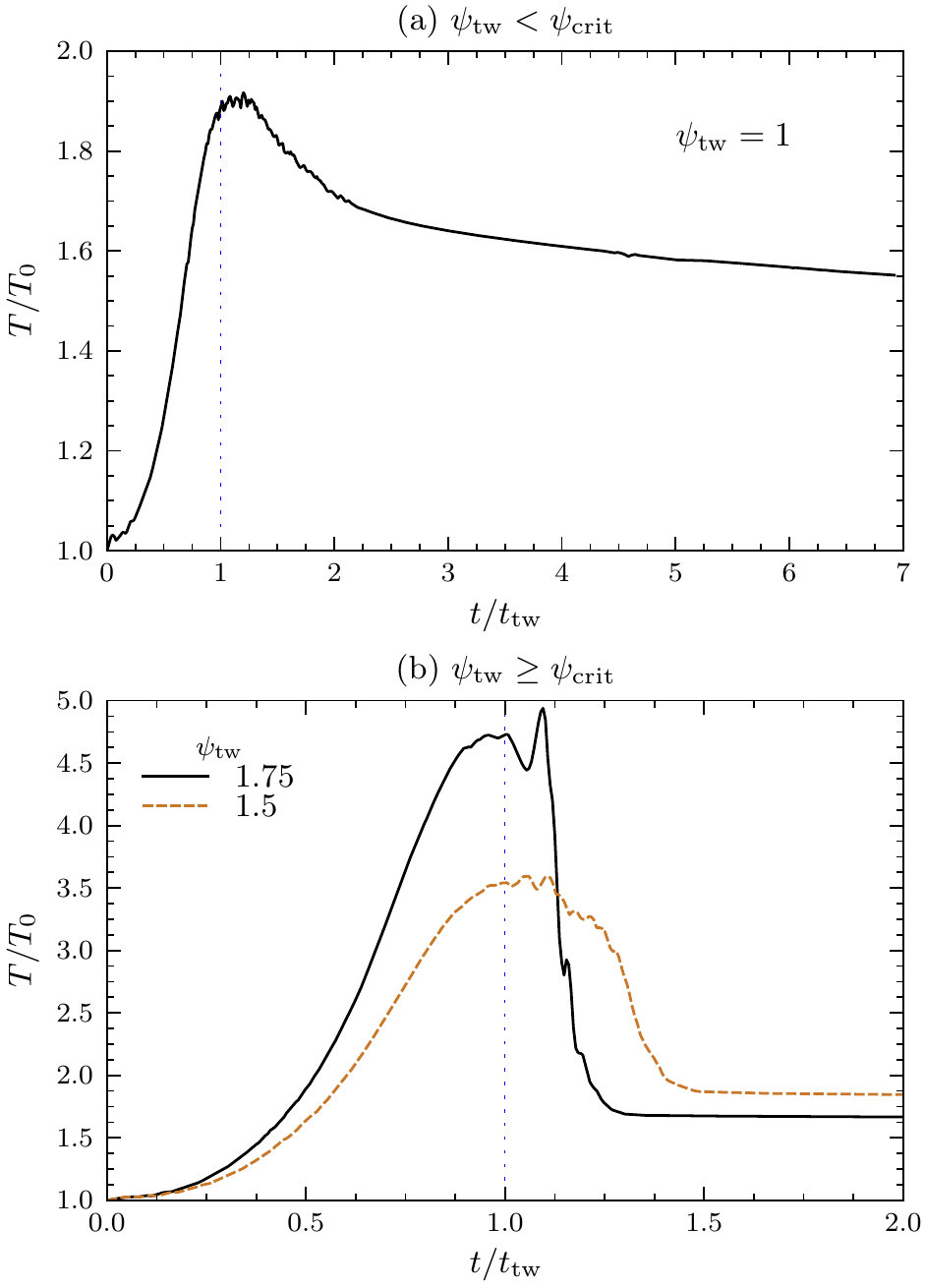}
\caption{ Torque on the star, in units of the initial torque due to rotation, for polar caps sheared to $\psi_{\rm tw}$ over time $t_{\rm tw}$. (a) $\psi_{\rm tw} < \psicrit$: slow diffusion of magnetic flux back through the current sheet; (b) $\psi_{\rm tw} \gtrsim \psicrit$: the magnetosphere undergoes large-scale reconnection to a twisted rotating steady state. A dotted line indicates $t = t_{\rm tw}$. \label{fig:rot-torque-crit}}
\end{figure}

\begin{figure*}[tp]
\centering
\includegraphics[width=165mm]{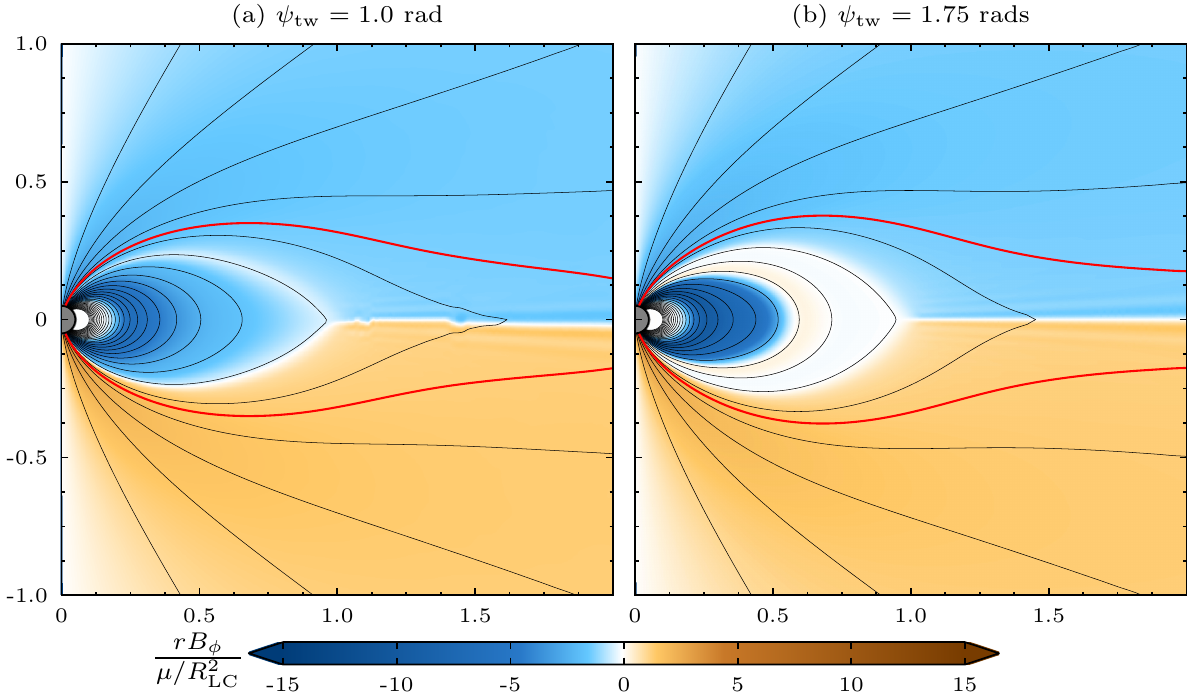}
\caption{Twisted rotating solutions at $t = 2\, t_{\rm tw}$, where $t_{\rm tw}$ is the time taken to implant a twist of amplitude $\psi_{\rm tw}$. (a)  $\psi_{\rm tw} = 1 < \psicrit$: a zone of untwisted flux slowly grows just inside the light cylinder, as some open flux diffuses back through the current sheet; (b) $\psi_{\rm tw} = 1.75 > \psicrit$: steady state following sudden reconnection of overtwisted flux. The bundle of twisted flux (with large $B_\phi$) inside the light cylinder is the twisted reservoir; the flux with $B_\phi \sim 0$ is the cavity. Color and field lines are as in Figure~\ref{fig:rot-early}. \label{fig:rot-pseudo-crit}}
\end{figure*}

In the $\psi_{\rm tw} = 1.0$ simulation, $t_{\rm tw}$ is reached without any large-scale dynamic motion taking place. For $t > t_{\rm tw}$, some of the field slowly diffuses back through the light cylinder due to dissipation in the current sheet. As the open flux decreases, so does the spindown torque, which fluctuates as plasmoids are created and expelled (Figure~\ref{fig:rot-torque-crit}a). Over time, a narrow band of approximately untwisted flux is formed in the closed zone, between the light cylinder and the twisted field lines which were never opened (Figure~\ref{fig:rot-pseudo-crit}a).

In the $\psi_{\rm tw} = 1.75$ simulation, the configuration gradually becomes less stable as the twisting rate is reduced to zero, as the dynamical stabilization effect due to the finite shearing rate (Section~\ref{sec:shearrate}) becomes weaker. At some point near $t_{\rm tw}$, a bundle of closed flux expands rapidly through the light cylinder, but the field lines beneath it are still in approximate pressure equilibrium and remain closed.
The inflation of this last opening field line is shown in Figure~\ref{fig:rot-reconn}: the field line with $u = 0.1435$ expands rapidly through the light cylinder, while the next field line shown (with $u= 0.1465$) does not. A current layer forms behind the $u=0.1435$ field line, which becomes thinner as its apex moves outward at approximately the speed of light---this is an ideal (dissipationless) collapse to a current sheet. Note that this collapsing current layer is distinct from the discontinuous current sheet just below it; the former comprises twisted field lines having $B_\phi$ of only one sign (here, negative) along their entire lengths, while the latter, created by rotation, is a discontinuity across which $B_\phi$ changes sign.

An X-point (or pinch) geometry is created at the cusp separating closed and inflating flux at $r \approx 0.8\, \rlc$, and the field gradients (and hence effective numerical resistivity) become large enough to trigger reconnection at the X-point. Reconnection first occurs across the twisted current layer, not the rotationally induced current sheet. Following reconnection, those parts of field lines which are attached to the star snap back toward it, removing pressure support around the X-point; this causes more flux to dive into the reconnection region, including field lines which previously straddled the current sheet (such as the $u =$ 0.125 and 0.135 field lines in Figure~{\ref{fig:rot-reconn}).  The O-geometry field line sections are untethered from the stellar surface, and are expelled, carrying most of the twist energy out of the system. The torque on the star plummets as more open flux becomes closed inside the light cylinder (Figure~\ref{fig:rot-torque-crit}b).  The reconnection timescale is much shorter than the field opening timescale. Field lines reconnect so as to become untwisted, which removes toroidal magnetic pressure support---the overextended field lines then snap back toward the star under high magnetic tension. In this way the reconnected untwisted field lines bury the underlying twisted ones, forcing them inward.

\begin{figure}[tp]
\centering
\includegraphics[width=85mm]{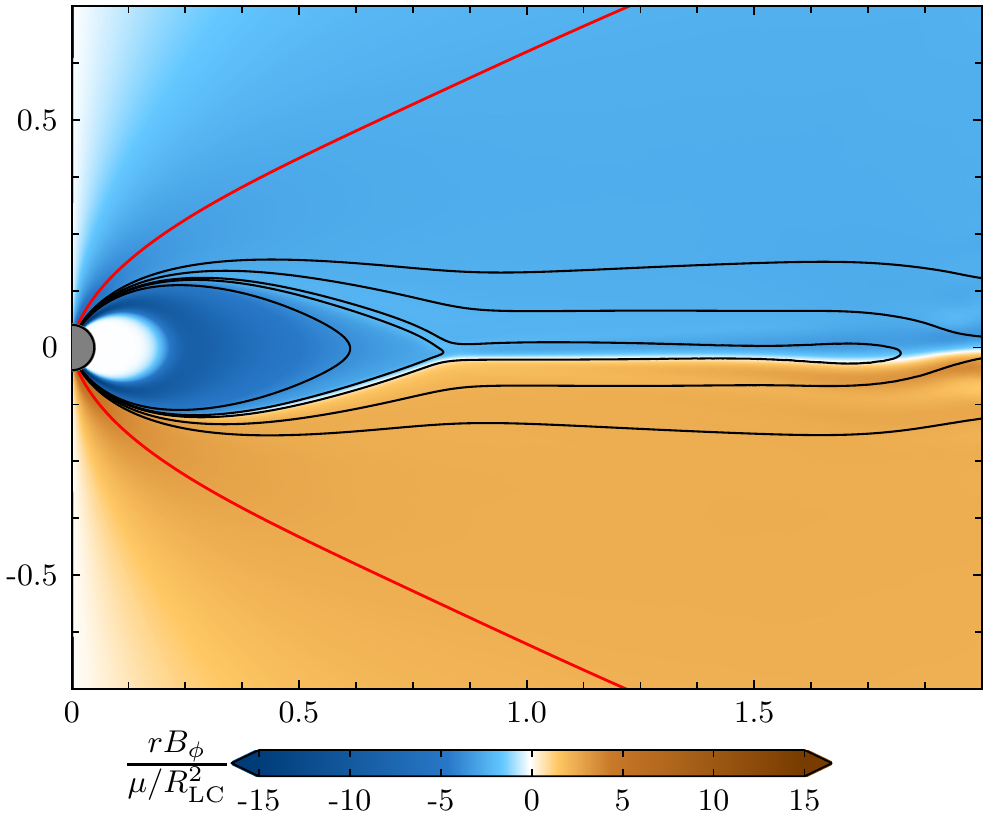}
\caption{Formation of the reconnecting current layer in the $\psi_{\rm tw} = 1.75$ simulation, at $t = 1.09\,t_{\rm tw}$. Black curves indicate field lines at $u =$ 0.125 (outermost line), 0.135, 0.1435, 0.1465, and 0.155 (innermost line); the red curve shows the field line that closes at the light cylinder at $t=0$. Reconnection will begin in the current layer between the $u =$ 0.1435 and 0.1465 field lines, which lies above the discontinuous current sheet (indicated by the color discontinuity). Color represents toroidal magnetic field multiplied by the radial coordinate. Axes are labeled in units of $\rlc$.  \label{fig:rot-reconn}}
\end{figure}

Reconnection continues until a new equilibrium is reached (Figure~\ref{fig:rot-pseudo-crit}b); since no further shearing is applied, this is a true equilibrium state, in contrast to the quasi-steady state in Figure~\ref{fig:rot-early}b. In the equilibrium solution, there is a bundle of strongly twisted field lines, the ``twisted reservoir,'' which were never open and so did not experience reconnection. There is also a region of untwisted closed flux, the ``cavity,'' between the reservoir and the flux surface closing at the light cylinder; the twist on these field lines was emitted to infinity while the field lines were open. The twist in the reservoir causes more flux to be open than in the initial untwisted state at $t=0$. In general, the larger the twist amplitude $\psi_{\rm tw}$, the more flux is opened and reconnects, and so the smaller the twisted reservoir (and the closer the new equilibrium state is to the untwisted rotating solution; Figure~\ref{fig:rot-torque-crit}b). 

We find that a slow mode of relaxation, due to ejection of small plasmoids, operates if $\psi_{\rm tw} = 1$; after $t_{\rm tw}$ a new equilibrium is gradually approached without  large-scale reconnection or significant dynamics. If $\psi_{\rm tw} = 1.75$ a catastrophic reconnection event occurs at $t_{\rm tw}$, and the configuration reaches a new equilibrium state on a much shorter timescale. The simulation with $\psi_{\rm tw} = 1.5$ displays intermediate behavior: first several smaller plasmoids are released, but the system soon enters an unstable rapid-reconnection phase, quickly reducing the spindown torque applied to the star (see Figure~\ref{fig:rot-torque-crit}b). This division into gradual relaxation and large-scale fast reconnection leads us to conclude that this system displays the same critical behavior as non-rotating twisted magnetospheres: there is a dramatic reconfiguration when the system is overtwisted, which occurs when the twist amplitude exceeds a critical value $\psicrit$. In the slow shearing limit, magnetic reconnection immediately follows the loss of equilibrium; a finite shearing rate can delay the onset of reconnection. The particular model we describe here has $\psicrit \approx 1.5$, which is smaller than the value for the corresponding non-rotating problem, $\psicrit \approx 3$ (\S~\ref{sec:shearrate}).

%%%%%%%%%%%%%%%%%%%

\subsection{Shearing through multiple reconnection events}

In the preceding section we described simulations in which a twist of a specific amplitude is implanted in the magnetosphere, and no shearing is applied thereafter. Now we consider models in which the shearing rate, rather than being returned to zero, is maintained at a constant value $\omega_0$, thus shearing the magnetic footpoints through large angles $\omega_0 t$ over the duration of the simulations. In the following we will use $\omega_0 t$ to label the applied surface shear instead of $\psi$, which we reserve for the angular displacement between the footpoints of a given field line (this will not be the same for all twisted field lines once the magnetosphere has experienced a reconnection event).

We begin with a model closely related to those described above: a polar cap of size $a=3$, twisted at a constant rate of $\omega_0 = \Omega/50$; the light cylinder is again at $\rlc=20$. The evolution of the torque $T$ in this simulation is shown by the red dotted curve in Figure~\ref{fig:rot-torque-omega}. At this constant shearing rate the magnetosphere moves smoothly through the critical twist amplitude $\psicrit \approx 1.5$, due to the dynamical stabilization effect (Section~\ref{sec:shearrate}). At $\omega_0 t \approx 3$, the timescale for inflating flux to collapse to a thin current layer becomes shorter than the timescale for rapid expansion of sufficient additional flux to prevent collapse, and a thin current layer forms (as illustrated in Figure~\ref{fig:rot-reconn}), initiating the first large-scale reconnection event.  Subsequently the field lines that have reconnected are untwisted, while those that did not reconnect remain strongly sheared---a twisted reservoir (Section~\ref{sec:rot-crit}). The stellar surface shearing continues, pumping ever more twist energy into the twisted reservoir. At $\omega_0 t \approx 4$ the twisted reservoir loses equilibrium and inflates outward rapidly in an explosive ``ring-like'' event, creating a sudden spike in spindown torque. From this we learn that rotating magnetospheres with twisted polar caps can experience ring-like second inflation phases, just like non-rotating magnetospheres (Section~\ref{sec:pcshear}); we also learn that these more explosive second events, caused by the magnetic detonation of the twisted reservoir, are associated with brief spikes in spindown torque. See \citet{2012arXiv1201.3635P} for a detailed description of the evolution of the twisting, rotating polar cap models through many expansion and reconnection events.

In \S~\ref{sec:shearrate} we described how the surface shearing rate strongly affects the twist amplitude at which the magnetosphere undergoes reconnection. We now investigate the effect of varying the rate at which a rotating magnetosphere is sheared. The spindown torque $T$, for simulations having the same sheared polar cap ($a=3$) but varying $\omega_0/\Omega$, is plotted against applied surface shear in Figure~\ref{fig:rot-torque-omega}. 

\begin{figure}[tp]
\centering
\includegraphics[width=85mm]{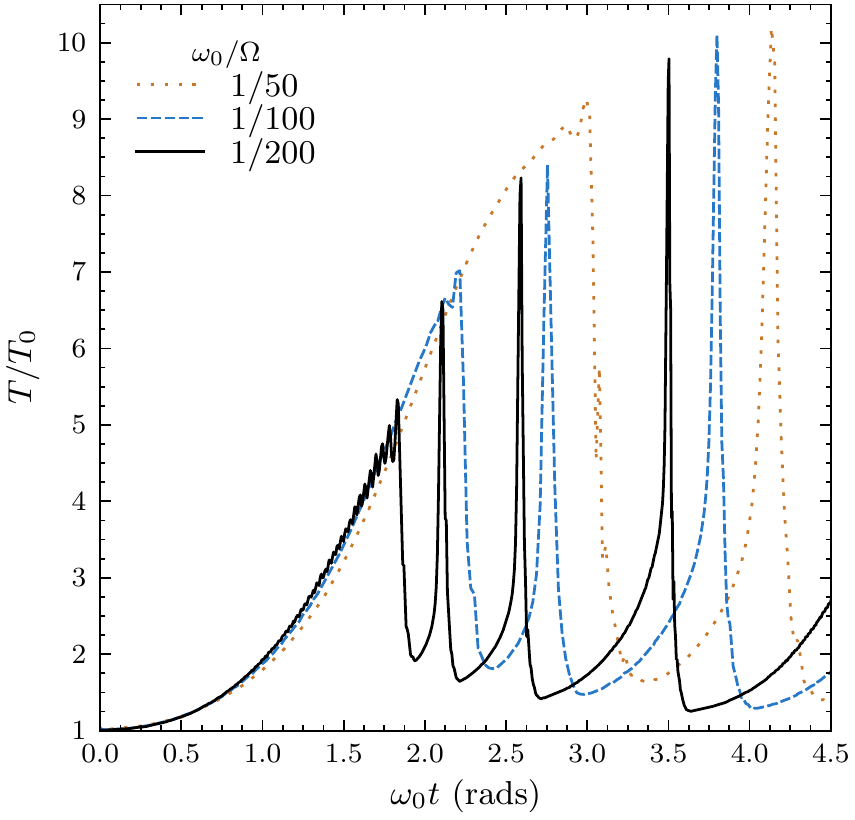}
\caption{ Torque on the star, in units of the torque $T_0$ (normal untwisted rotator at $t=0$), for a polar cap with $a=3$ and varying twisting rate $\omega_0$. \label{fig:rot-torque-omega}}
\end{figure}

Up to $\omega_0 t \sim 1.5$, the torque is very similar for the three runs shown in Figure~\ref{fig:rot-torque-omega}. At this point the run with lowest shearing rate, $\Omega/200$, enters a phase of magnetospheric ``breathing,'' where bundles of closed flux inflate rapidly through the light cylinder, are initially unsupported by expanding underlying flux and so begin to close again, but are prevented from initiating runaway large-scale reconnection by the eventual expansion of the lower-lying field. These increasingly vigorous motions are accompanied by oscillations in $T$ of growing amplitude and period $\sim 7 \rlc/c$. At $\omega_0 t \approx 1.8$ the newly opened field lines begin to collapse on a timescale shorter than the time needed for the lower-lying flux to push them back out through the light cylinder. Then runaway reconnection begins. It creates a zone of untwisted closed field lines inside the light cylinder; these field lines confine the inner twisted reservoir. A large fraction of the flux opened by twisting reconnects and returns to the closed zone with zero twist. 

Faster shearing can stabilize the magnetosphere against large-scale reconnection, similar to the results of simulations without solid-body rotation $\Omega = 0$; in these simulations with larger $\omega_0/\Omega$ the breathing phase is absent. However, the twist amplitude needed to move between subsequent brief ring-like explosive events is less sensitive to the shearing rate, because these are due to sudden unstable dynamic expansion of the twisted reservoir, and the overlying less-twisted flux acts like an elastic nozzle, inducing reconnection soon after inflation (see Section~\ref{sec:pcshear}).

Both the smooth torque growth and the sudden torque spikes in Figure~\ref{fig:rot-torque-omega} lie on a common envelope, which is independent of the shearing rate. This implies that the maximum possible torque enhancement at any time $t$ depends only on the twist $\omega_0 t$ which has accumulated on the most twisted field lines, irrespective of whether twist has been removed from other field lines in previous reconnection events. The peak heights increase with $\omega_0 t$ as the explosive inflation is driven by ever deeper field lines, opening more magnetic flux through the light cylinder.

For $\omega_0 = \Omega/200$, $\psirec$ is roughly equal to the critical twist amplitude estimated in the preceding section. This shearing rate is used in the simulations described below.

Increasing the size of the twisted polar cap allows a larger multiple of the star's rotationally opened flux to be pushed through the light cylinder, giving a greater spindown torque enhancement. We have run simulations with $a = $ 2 to 12 with shearing rate $\omega_0 = \Omega/200$, and find that the peak spindown torque scales as, and for the largest events is roughly equal to, $a^2$,
\beq
\frac{T_{\rm peak}}{T_0} \approx a^2.
\label{eq:T_scale_a}
\eeq
The torque is plotted against applied shear in Figure~\ref{fig:rot-torque-a}, where the quadratic scaling of the spike heights is apparent. The evolution to the maximum footpoint displacement shown ($\omega_0 t = 8$) takes roughly 263 stellar rotation periods with shearing rate $\omega_0 = \Omega/200$.  Note also that solutions with larger $a$ undergo their first reconnection event at larger twist amplitude, since twisted deep flux pushes the overlying field lines outward; this slightly greater expansion allows the dynamical stabilization mechanism to be effective to larger twist. The average torque enhancement over the length $t_{\rm tot}$ of the simulation, 
\beq
\left< \frac{T}{T_0} \right> = \frac{1}{t_{\rm tot} T_0}\int_0^{t_{\rm tot}} T(t')\,{\rm d}t',
\label{eq:Tavg} 
\eeq
is 2.64 for $a=3$, 5.45 for $a=6$, and 10.75 for $a=12$ (where $\omega_0 t_{\rm tot} = 8$). In our simulations of twisted polar caps, $\left<T/T_0\right> \sim a$ when averaged over several flux expansion and reconnection events.

\begin{figure}[tp]
\centering
\includegraphics[width=85mm]{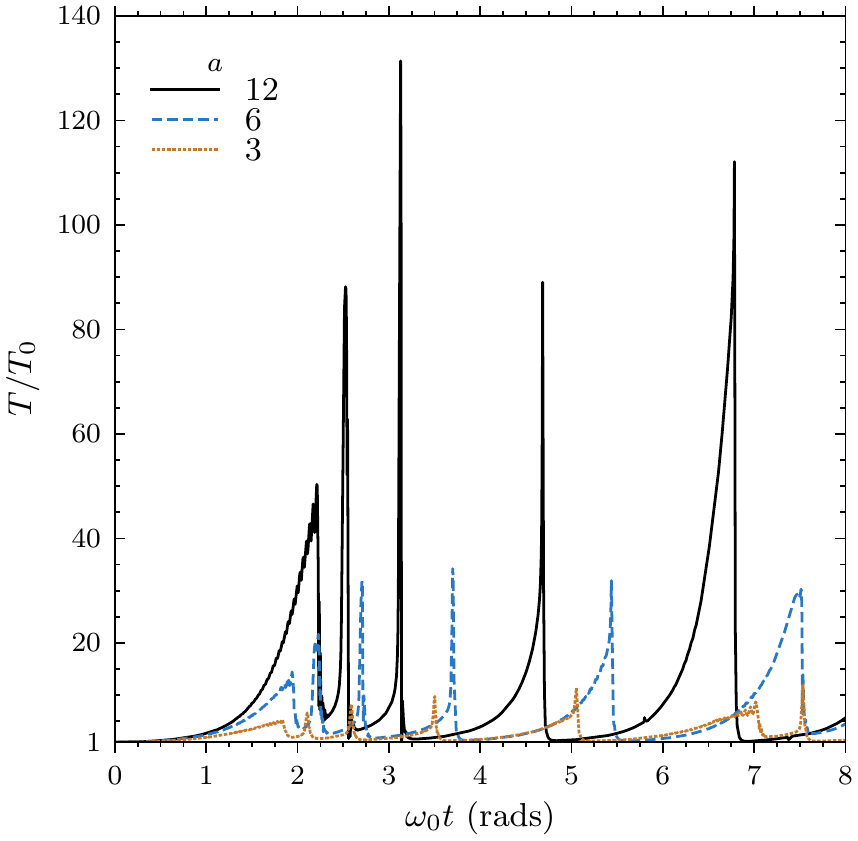}
\caption{Torque on the star for polar caps of varying sizes $a$, with shearing rate $\omega_0 = \Omega/200$. The red dotted curve in this figure represents the same simulation as the solid black curve in Figure~\ref{fig:rot-torque-omega}. \label{fig:rot-torque-a}}
\end{figure}

\subsection{Ring shearing in the closed magnetosphere}

The behavior seen in Figure~\ref{fig:rot-torque-a} for the polar cap shearing model is complicated---expansion and reconnection events can be gradual (on the twisting timescale) or explosive (on the light-cylinder-crossing timescale), or lie somewhere between the two, depending on the how the magnetosphere has evolved up to the time of the event. The twist evolution for a narrow ring shear profile (Figure~\ref{fig:profiles}c) is simpler. In this case, there is little poloidal expansion (and hence increase in torque) until a large twist amplitude has accumulated, at which point the twisted flux bundle inflates explosively under high magnetic pressure. Figure~\ref{fig:rot-torque-ring} shows the torque enhancement for two ring models having the same ratio of twisted to rotationally opened flux, but located at different latitudes. Both simulations have, on average, narrower torque spikes than the polar cap models (Figure~\ref{fig:rot-torque-a}), with the more deeply buried twisted ring powering distinctly more explosive events (only two in eight radians of twisting, and both on the light-cylinder-crossing timescale). The lower flux surface of the ring, $a_2$, determines the torque peak height, because during an eruption the twisted field line bundle pushes some of its own flux and all of the overlying flux (at smaller $u$) through the light cylinder, 
\beq
\frac{T_{\rm peak}}{T_0} \approx a_2^2.
\label{eq:T_scale_a_2}
\eeq
Opening of more deeply buried flux (i.e. further from the open flux bundle), requires a larger twist amplitude. This leads to a more powerful detonation, which releases more magnetic energy over a shorter time.

\begin{figure}[tp]
\centering
\includegraphics[width=85mm]{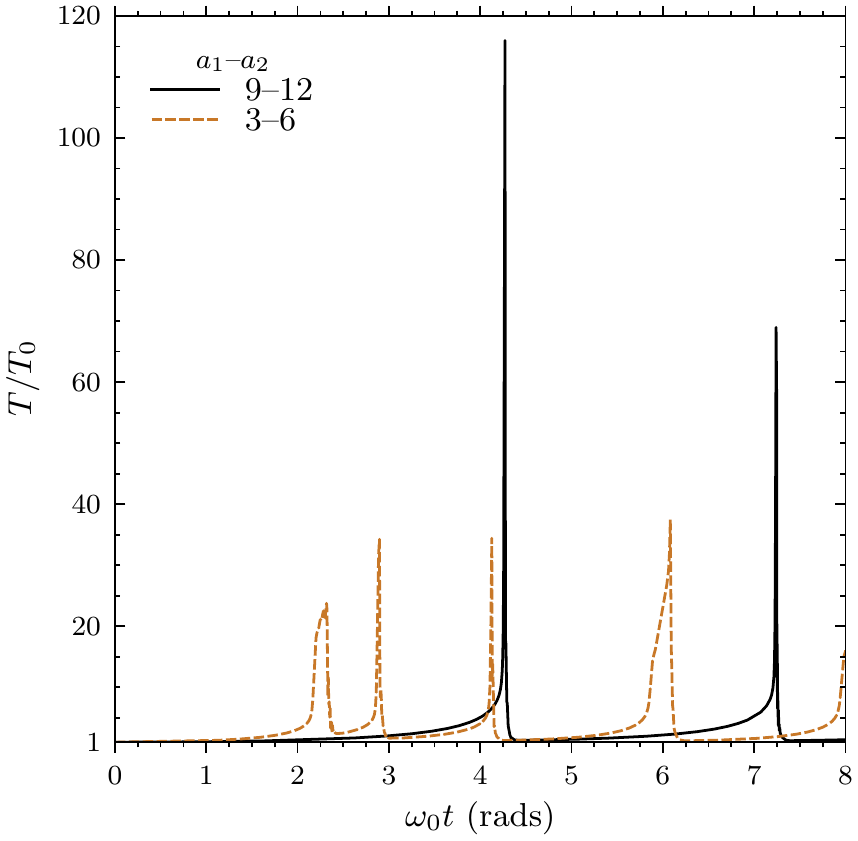}
\caption{Torque enhancement due to slow twisting in two ring shearing models, having the same amount of twisted flux but located at different latitudes. \label{fig:rot-torque-ring}}
\end{figure}

The first detonation in the $a=$ 9--12 simulation is shown in Figure~\ref{fig:rot-pseudo-ring}. At small to moderate applied shear, the deeply buried twisted flux bundle aquired a large toroidal field component, but there is little poloidal expansion---even at $\omega_0 t=2$, only a small amount of additional flux has been pushed through the light cylinder (Figure~\ref{fig:rot-pseudo-ring}a). When the shear is close to the critical twist angle for the twisted flux, the twisted bundle begins to expand poloidally, causing more of the overlying field lines to open; in this phase the magnetosphere becomes more sensitive to the twist amplitude, but the field expands stably on the shearing timescale (Figure~\ref{fig:rot-pseudo-ring}b). At the critical twist amplitude, the twisted flux bundle loses equilibrium, and inflates rapidly under its high magnetic pressure (Figure~\ref{fig:rot-pseudo-ring}c). This unstable inflation occurs on a dynamical timescale, on the order of $\rlc/c$. Near the base of the expanding flux, a current sheet forms behind the last expanding field line; in Figure~\ref{fig:rot-pseudo-ring}c, this current sheet will be just below the field line drawn at $u=0.4$. Catastrophic unstable reconnection then takes place in the current sheet, as described in \S~\ref{sec:rot-crit}, leaving a smaller twisted reservoir. The Y-point at the boundary of the newly formed cavity retreats back out to the light cylinder during the reconnection process (Figure~\ref{fig:rot-pseudo-ring}d). The magnetosphere returns to an equilibrium spindown steady state after the Y-point reaches the light cylinder.

\begin{figure*}[tp]
\centering
\includegraphics[width=165mm]{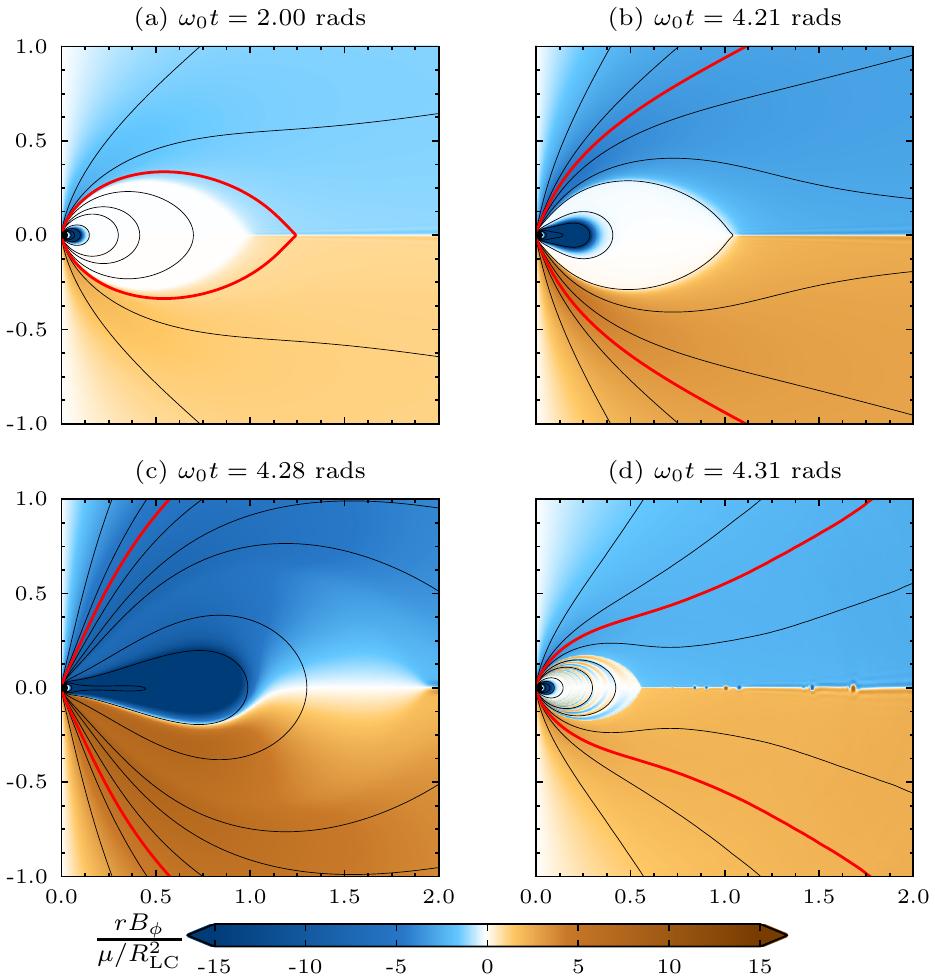}
\caption{The first explosive reconnection event for a twisted ring with $a=9$--12. Color is as in Figure~\ref{fig:rot-early}; the black curves show poloidal field lines, one drawn at $u=0.01$, four equally spaced between $u=$ 0.025 and 0.1, and five between $u=$ 0.1 and 0.5. The red curve indicates the field line initially closing at the light cylinder. Axes are labeled in units of $\rlc = 40\,\rstar$. \label{fig:rot-pseudo-ring}}
\end{figure*}

The widths of the various torque spikes for each of the five simulations are given in Table~2, where their full widths at half maximum (FWHM) are scaled to the light-crossing time of the light cylinder, $\rlc/c$. The polar cap models ($a = $ 3, 6, 12) have several kinds of events: rapid explosions on roughly the light-crossing time, gradual inflations on the twisting timescale $\gg \rlc/c$, and those which are somewhere between the two. The ring simulations generally have briefer, more dramatic, events---the $a =$ 9--12 run in particular has only two torque spikes, both of  duration $\sim 2\rlc/c$. 

\begin{deluxetable*}{l c c c c c c}
\tablewidth{110mm}
\tablecaption{Torque peak widths: $\Delta t_{\rm FWHM}/(\rlc/c)$}
\tablehead{\colhead{$a$} & \colhead{$\rlc$} & \colhead{Peak 1} & \colhead{Peak 2} & \colhead{Peak 3} & \colhead{Peak 4} & \colhead{Peak 5}}
\startdata
3 & 20 & 79.5 & 8.4 & 7.5 & 6.7 & 12.3\\
6 & 20 & 65.3 & 15.2 & 6.4 & 4.4 & 17.9 \\
12 & 40 & 55.7 & 11.9 & 4.5 & 2.8 & 30.6 \\
3--6 & 20 & 29.9 & 5.9 & 3.1 & 25.9 & 56.9\\
9--12 & 40 & 2.3 & 2.1 &      &          &
\enddata
\end{deluxetable*}

Magnetic energy is dissipated into thermal energy during the reconnection phase, which corresponds to the sharp downstrokes in the torque curves in Figures~\ref{fig:rot-torque-omega} to \ref{fig:rot-torque-ring}. In the simulations we have performed, the reconnection timescale $\Delta t_{\rm rec}$ following gradual ``polar-cap-like'' expansion is $\Delta t_{\rm rec} \approx$ 100--200 $\rstar/c$, while after explosive ``ring-like'' flux breakout the timescale is shorter, $\Delta t_{\rm rec} \approx$ 50--70 $\rstar/c$; scaled to the radius of a neutron star ($\rstar \approx 10^6$~cm), these results correspond to reconnection lasting roughly 1.5 to 6 ms. As in the non-rotating simulations, the reconnection timescale is approximately $\Delta t_{\rm rec} \sim 2 R_{\rm rec}/v_{\rm rec}$, where $R_{\rm rec}$ is the initial inner radius of the reconnecting current sheet and $v_{\rm rec} \sim 0.1 c$.

\subsection{Asymmetry, linear momentum transfer, and trapped Alfv\'{e}n waves}

We have discussed how twisting of the magnetosphere increases the spindown torque on the star. The rotating simulations are slightly asymmetric about the equator, since only northern latitudes are sheared, raising the possibility of a net transfer of linear momentum along the star's rotation axis, which we shall call the $z$-axis. A net force could arise from asymmetric spindown as the magnetosphere is inflated, or from asymmetry in the final dynamic phase in which the overtwisted flux is ejected.  

An upper bound can be placed on the net force by noting that the momentum density carried by the electromagnetic field does not exceed its energy density divided by $c$. Therefore the greatest momentum that can be transferred from the magnetosphere, if it is entirely emitted to infinity from one side, is $p_{\rm max} = W_{\rm max}/c$, where $W_{\rm max} \approx W_{\rm pot}$ is the maximum energy that can be stored, and so
\beq
p_{\rm max} \approx \frac{\mu^2}{3 \rstar^3 c}.
\eeq
The maximum estimated force, if this momentum is transfered over a time $\rstar/c$, is then 
\beq
F_{\rm max} \approx \frac{\mu^2}{3 \rstar^4} = 3\times 10^{41}\, \lp\frac{\mu}{10^{33}\,{\rm G\, cm^3}}\rp^2\, \lp\frac{10^6\, {\rm cm}}{\rstar}\rp^4 \, {\rm dyn.}
\label{eq:fmax}
\eeq
For a neutron star of mass $M_* = 1.4\, M_{\odot}$, momentum conservation $M_* v_{\rm kick} = p_{\rm max}$ implies a maximum kick velocity of only $v_{\rm kick} \sim 0.03$ km s$^{-1}$, for an ultra-strong magnetic field $B \sim 10^{15}$ G that corresponds to $\mu \sim 10^{33}$ G cm$^3$.

Even though a dramatic rocket effect is impossible, it is interesting to calculate the force applied along the $z$-axis, as it provides a measure of the asymmetry of the inner magnetosphere during a flare-like eruption. The flux of $z$-momentum onto the $r = \rstar$ surface is given by the $T_{rz}$ component of the electromagnetic stress-energy tensor, which can be integrated over the stellar surface to give the total instantaneous force on the star. 

The net force during the first explosive event of the $a=$ 9--12 ring simulation (as illustrated in Figure~\ref{fig:rot-pseudo-ring}) is shown in Figure~\ref{fig:zforce}. The torque spike for this event has a full width at half maximum of $\Delta t = 92\,\rstar/c$ (Table~1). Before reconnection begins, there is a small force pushing the star in the negative $z$-direction, which increases on the same timescale as the torque, forming a peak of similar FWHM and with peak value $-6\times 10^{-6}\, F_{\rm max}$. This small force implies that the magnetosphere is very nearly symmetrical about the equator even during the explosive inflation phase.

\begin{figure}[tp]
\centering
\includegraphics[width=85mm]{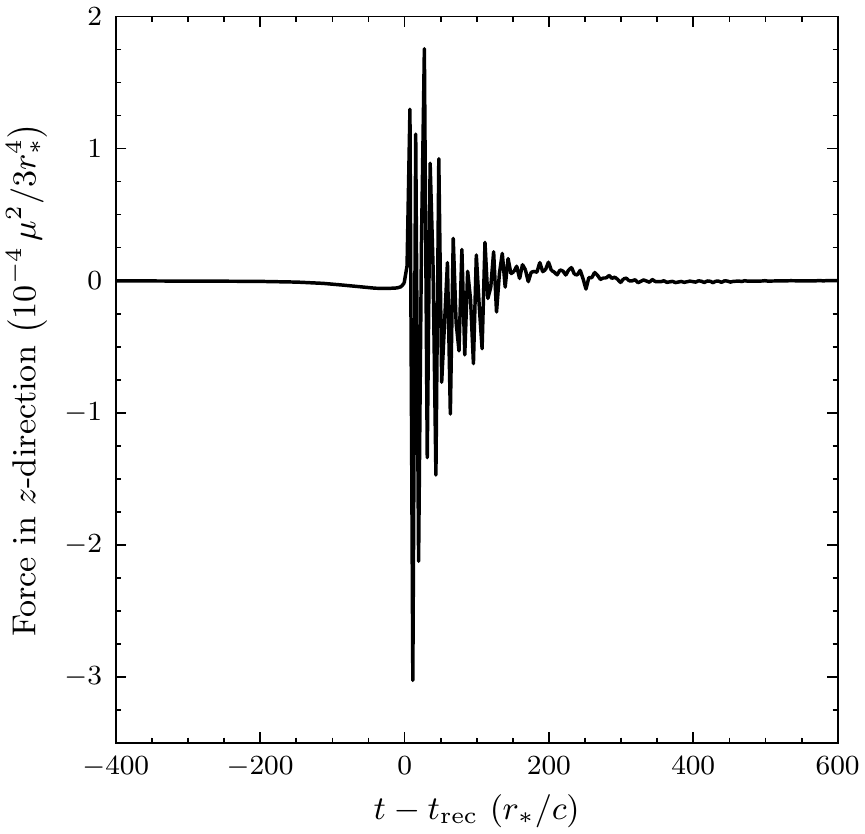}
\caption{Force along the rotation axis during a flare-like event, where $t_{\rm rec}$ indicates the onset of reconnection and the force is scaled to $10^{-4}$ times the estimated maximum force, $F_{\rm max}$ (Equation~(\ref{eq:fmax})). \label{fig:zforce}}
\end{figure}

The reconnection process injects Alfv\'{e}n waves onto the newly closed field lines. These waves, clearly visible in our simulations, travel from the reconnection region to the stellar surface, are reflected, and continue to bounce backward and forward on the closed lines; at the same time, reconnection continues to inject new waves, and the closed flux bundle oscillates as it adjusts to a new pressure equilibrium state. The field lines are not exactly symmetrical, and so waves from the reconnection region (and subsequent reflected waves) are incident on the surface at slightly different times, giving the star small net kicks of alternating sign along the axis. These impulses are responsible for the rapidly oscillating net force after $t_{\rm rec}$ seen in Figure~\ref{fig:zforce}, which can be significantly larger than the peak force before reconnection. The trends on longer timescales may be due to the equilibration of the closed flux. The small-scale Alfv\'{e}n waves are gradually removed by numerical dissipation, and are gone by roughly $t = t_{\rm rec} + 500 \,\rstar/c$.

%\clearpage

\section{Discussion}
\label{sec:discussion}

Using numerical simulations, we have studied the response of relativistic magnetospheres to slow twisting of the magnetic field lines, driven by shearing of the stellar surface into which the lines are frozen. Although the general force-free twisting problem has relevance to many astrophysical objects, including accretion discs and gamma-ray bursts, here we focus on applications to magnetar theory. 

The magnetospheric twisting model may explain two key aspects of magnetar activity. First, twisting injects current-carrying Alfv\'{e}n waves into the magnetosphere, setting up a system of large-scale magnetospheric currents. The particles which constitute these currents produce magnetars' persistent non-thermal emission \citep{2002ApJ...574..332T, 2013ApJ...762...13B}. Second, magnetic configurations can become unstable when strongly sheared (``overtwisted''). The ensuing dynamic evolution results in reconnection and the release of magnetic energy, a possible mechanism for magnetar bursts and giant flares \citep{1995MNRAS.275..255T, 2006MNRAS.367.1594L}. Our simulations demonstrate what happens when an axisymmetric relativistic magnetosphere is twisted beyond the instability threshold.

\subsection{Equilibrium solutions}

When the shearing is applied slowly and the twist amplitude is modest, the magnetosphere moves quasi-statically through a sequence of equilibrium states. As the twist amplitude $\psi$ increases, these equilibria have increasing toroidal magnetic field, current density, and total magnetic energy. The field lines in the poloidal plane expand outward under the increased magnetic pressure from the toroidal field. The expansion rate, ${\rm d}R_{\rm max}/{\rm d}\psi$, is negligible at $\psi=0$ and increases with $\psi$ (Figure~\ref{fig:rmax}). Eventually many field lines inflate so much that the toroidal field at each point along them decreases even as the total integrated twist angle continues to grow. In the theoretical final state in the equilibrium sequence, all sheared field lines are open to infinity (i.e. have infinite $R_{\rm max}$), there is no toroidal magnetic field, and the current density is confined to discontinuous current sheets. 

The particular final equilibrium state in which the entire stellar surface is sheared has the maximum energy of any configuration with the same surface normal magnetic field distribution; for the dipole distribution we use, this maximum energy is $W_{\rm max} = 1.662\, W_0$, where $W_0$ is the energy of the untwisted dipole field. This is the greatest magnetic energy that can be stably stored in the magnetosphere. Our numerical solutions clearly evolve toward this state when shearing is applied down to the equator. We produce a configuration that has energy $W = 1.619\, W_0$, energetically 94\% of the way from the initial dipole field to the end point of the equilibrium sequence.

The energy stored in the magnetosphere is less if the twisting is not applied globally. For example, a twisted polar cap encompassing half of the star's total magnetic flux can only store approximately $0.07\, W_0$ in free energy and remain in equilibrium (Figure~\ref{fig:energy_equil}). This reduces the size of the energy reservoir which can be released in a subsequent explosion. More energy can be stored if the twisting is confined to a ring of deeply buried field lines, because the tension in the overlying untwisted flux has a confining effect, preserving equilibrium to larger twist amplitudes. One can estimate the magnetic free energy using the approximation that poloidal field expansion is negligible, Equation~(\ref{eq:Westimate}); we find that this estimate is accurate for $\psi \lesssim 1.3$,  and overpredicts $W$ for $\psi > 1.3$.

\subsection{Critical twist amplitude, overtwisting, and reconnection}

It has long been argued that at some critical point the twisted magnetosphere would lose equilibrium and enter a fully dynamic state, possibly resulting in magnetic reconnection and the dissipative release of energy. Our simulations support this view. For every shearing profile on the star $\omega(\theta)$, there appears to be a critical twist amplitude $\psicrit$. Configurations with $\psi < \psicrit$ are indefinitely stable, while no stable state having $\psi > \psicrit$ can be created. When the magnetosphere loses equilibrium, field lines inflate at the speed of light toward the open state, forming a thin current layer which collapses into a discontinuous current sheet. 

The collapse initiates a phase of unstable reconnection, and all of the flux opened by twisting reconnects to form a fully closed magnetosphere, as illustrated in Figure~\ref{fig:polar-recon}. In the process some of the stored magnetic energy is expelled as large-scale Poynting flux and some is dissipated into heat. The current layer collapse and subsequent plasmoid-generating reconnection appear to be due to the tearing instability. During reconnection, magnetic flux moves toward the current sheet at $v_{\rm rec} \sim 0.1\, c$ and newly reconnected field lines leave the reconnection region at nearly the speed of light. In general we find that the duration of the reconnection phase is related to the initial inner radius $R_{\rm rec}$ of the reconnecting current sheet: $\Delta t_{\rm rec} \sim 2 R_{\rm rec}/v_{\rm rec}$. In our simulations, resistivity is applied numerically where field gradients are strong, and so we have not self-consistently calculated the reconnection speed, which may depend on the microphysics of the reconnecting current layer. The magnetic field is driven into the reconnection region by global magnetic stresses and leaves it at almost the speed of light; therefore the reconnection rate may turn out to be insensitive to the resistive length scale and the detailed operation of the resistive processes.

We find that the fraction of the free energy of the twist that is dissipated in the dynamic phase is about 15\% for the equatorial shearing model (Section~\ref{sec:eqshear}) and nearly 44\% in our fiducial polar cap simulation (Section~\ref{sec:pcshear}). While the dissipation fraction appears to be responsive to some numerical parameters that control resistivity in the current sheets, these large fractions are encouraging if one wishes to power energetic gamma-ray flares with liberated twist energy.

Finding the critical twist amplitude is complicated by an additional effect, which is that the continual twisting is itself temporarily stabilizing (Section~\ref{sec:shearrate}). Reconnection is inevitable when $\psi$ exceeds $\psicrit$, but its onset can be delayed, allowing $\psi$ to temporarily exceed $\psicrit$; the faster the twisting, the larger the twist amplitude which can be reached before reconnection begins. A larger twist amplitude translates into more energy stored in the magnetosphere and potentially more energetic bursts or flares when the energy is released. This is analogous to riding a bicycle---one is more stable against falling over when one pedals more quickly, while also storing more kinetic energy which can be released in a collision. At twist amplitudes above $\psicrit$, the solution is no longer in a quasi-equilibrium state, because there is no corresponding equilibrium to which the solution would gently relax (without change of magnetic topology) if the shearing were halted; the solution is rather in a ``dynamically stabilized'' state, entirely dependent for its stability on the maintenance of the shearing of the stellar surface.

The plastic motion of a magnetar's surface is expected to be slow, and so the dynamical stabilization would likely not be a large effect, leading in this case to instability and reconnection close to the critical twist angle. On the other hand, proto-magnetars and other highly energetic objects may have large, even relativistic, shearing rates; these systems may be able to maintain stability against reconnection up to larger twist angles, store more energy in their magnetospheres, and so produce more powerful flares during the subsequent reconnection phase. For example, a polar cap model whose critical twist angle is approximately 3 radians can be stabilized up to a twist angle of more than 12 radians by shearing at $\omega =$ 0.01 $c/\rstar$ (this may be an underestimate, because the realistic resistive length scale may be smaller than that in our simulations).

The resistive instability initiates the collapse of the thin current layer to an unresolved current sheet. The collapse dynamics are sensitive to the resistivity level, and we have found no clear evidence for an ideal or resistivity-independent unstable collapse phase. However, the critical twist angle at which collapse occurs is only weakly dependent on the effective resistivity (which in our simulations can be changed via either the grid scale or the spectral filtering level). This is because at large twist amplitudes the magnetosphere becomes extremely sensitive to further shearing, and so a small increase in applied shear quickly brings the current layer to the point of resistive instability, regardless of how small resistivity may be. It is in this sense that we speak of an effectively resistivity-independent critical twist amplitude, the end point of the sequence of equilibria constructed by arbitrarily slow twisting, determined only by the shearing profile on the star. These conclusions apply only to the axisymmetric configurations we have studied; in a fully three-dimensional scenario additional instability modes become available, possibly leading to ideal unstable field opening.

In simulations of a twisted polar cap, the magnetic field expands gradually and many field lines are  inflated to large distances (effectively becoming open) before reconnection. When a narrow ring of flux is twisted the expansion is delayed by the blanketing untwisted flux, and near the instability point the twisted flux bundle suddenly inflates violently in a ``magnetic detonation.'' The untwisted flux acts as an elastic nozzle, preventing the outflow from expanding easily in the meridional direction, and many field lines reconnect while their apexes are still relatively close to the star, $R_{\rm max} \lesssim 100 \rstar$. An explosive ring-like event can in fact be produced by a twisting polar cap, if the shearing is continued following the first reconnection event; in this case, some of the strongly twisted low-lying field lines, which did not reconnect in the first event, become unstable as they are twisted further, opening explosively before appreciable twist accumulates on the field lines above them.

\subsection{Rotating, twisting solutions}

Many features of the non-rotating twisting problem described above are also present when stellar solid-body rotation is added. There is still a critical twist amplitude beyond which the magnetosphere is unstable to catastrophic large-scale reconnection, and an approximate dichotomy between gradual, or polar cap-like, events and those which are explosive, or ring-like. Rotation changes the evolution in detail; for example, the relaxation of a rotating quasi-equilibrium configuration to a true stationary equilibrium (when the stellar shearing is withdrawn) involves some diffusion of open flux back through the light cylinder, and the critical twist amplitude is smaller in the rotating than in the non-rotating solutions.

Most importantly, rotation adds a new spatial scale, the light cylinder radius $\rlc$, and a new physical quantity, the spindown torque $T$ applied to the star by the magnetosphere. As in the non-rotating solutions, twisting causes expansion of the magnetic field in the poloidal plane; this pushes more flux through the light cylinder, increasing the spindown torque: $T/T_0 > 1$, where $T_0$ is the torque applied by the rotating untwisted magnetosphere. When twisting a polar cap, initially extending over both open and closed flux, the torque first increases on the twisting timescale $\omega^{-1}$; eventually,  the magnetosphere beyond the light cylinder becomes unstable to large-scale reconnection, as illustrated in Figure~\ref{fig:rot-reconn}. 

Reconnection of open flux buries a ``reservoir'' of strongly twisted field lines (e.g.\ Figure~\ref{fig:rot-pseudo-crit}b); as these continue to be twisted, they soon become unstable to explosive opening, resulting in narrow torque spikes of duration only a few times the light-crossing time of the light cylinder (Figure~\ref{fig:rot-torque-a}). The heights of these spikes are approximately $T/T_0  \approx a^2$, where $a$ is the ratio of the twisted flux to the flux opened by rotation alone. Only the brief, explosive events occur if, instead of a polar cap, a ring containing only closed flux is twisted (Figures~\ref{fig:rot-torque-ring} and \ref{fig:rot-pseudo-ring}).

Significant enhancement of the stellar spindown rate preceding a flare is a prediction of this model. The spindown rate should increase rapidly before the flare is observed. It is difficult to make specific predictions, because the timescale of the torque enhancement could be anywhere from the light-crossing timescale of the inner magnetosphere or light cylinder to the (possibly very long) twisting timescale, and depends on the shearing profile and the history of the system (i.e. whether the magnetosphere has recently produced a flare). In our simulations the briefest torque spikes lasted a few times the light cylinder light-crossing time, $R_{\rm LC}/c$ (Table 2). Magnetars rotate slowly (with periods of 2--12 s) and so have large light cylinders, of roughly $10^4$ stellar radii. It is entirely possible that a rapidly expanding unstable flux bundle would undergo reconnection before reaching the light cylinder, reducing the torque spike duration below $R_{\rm LC}/c$.

The most luminous flares require the participation of a large fraction of the magnetosphere; since magnetars spin relatively slowly, this translates into large values of the twisted-to-rotationally-opened flux ratio $a \gtrsim 10^3$, implying enormously accelerated spindown, on the order of $T/T_0 \gtrsim 10^6$, during the field expansion phase. 

This enhanced spindown rate will lead to a significant increase in period if sustained over a timescale of order the spin period or longer, as can occur when the twisting timescale is much longer than the period, $\omega \ll \Omega$. On the other hand, if twisting is fast, and a large twist amplitude is implanted in a fraction of a rotation period, the accelerated spindown phase will have a duration approximately the sum of the twist implantation and subsequent reconnection timescales. Since the reconnection timescale is likely to be short, the accumulated period increase will be much smaller than what is achievable in the slow-twisting case. We defer further discussion of magnetospheric dynamics in the fast-twisting regime to future work.

Our rotating solutions have $\rlc =$ 20--40 $\rstar$, corresponding to spin periods of $P \approx$ 4--8 ms, much shorter than those of magnetars. Slow rotation leaves the majority of field lines unchanged from the non-rotating dipole field, and the critical twist amplitudes will be similar to those found in Section~\ref{sec:shearprof}, as long as the flux surfaces bounding the twisted field lines are much larger than the rotationally opened flux ($a$, $a_1$, $a_2 \gg 1$). The critical twist amplitude will be reduced if all of the twisted flux is near the rotationally opened flux (Section~\ref{sec:rot-crit}). At lower rotation rates, expansion and reconnection events will continue to be divided into gradual and explosive categories, and be associated with torque spikes scaling quadratically with twisted flux, as given by Equations~(\ref{eq:T_scale_a}) and (\ref{eq:T_scale_a_2}).

This kind of torque enhancement due to dynamically opened magnetic flux may have been responsible for the ``braking glitch'' (or ``anti-glitch'') coincident with the 27 August 1998 giant flare of SGR~1900+14 \citep{1999ApJ...524L..55W}. Around the time of the flare (the precise timescale is unknown due to an 80 day gap in observations), the spin period of the source increased abruptly, with $\Delta P/P  = 10^{-4}$. This change is consistent with that produced by one of the explosive events described above, having sufficient twisted flux to provide the observed radiated energy and lasting about a light cylinder light-crossing time \citep{2012arXiv1201.3635P}. In contrast, an upper limit of $\Delta P/P \leq 5\times 10^{-6}$ was placed on the much more powerful giant flare from SGR~1806-20 \citep{2007ApJ...654..470W}.  This source may have experienced rapid twisting $\omega > \Omega$, or alternatively slow twisting and early reconnection of the unstable flux bundle, leading to a shorter phase of enhanced spindown torque and hence a smaller change in period. It is also possible that our axisymmetric model is insufficient to describe the variety of torque spikes introduced when the rotation, magnetic, and twisting axes are misaligned.

The spindown rate is predicted to increase most dramatically just before flares or bursts. However, twisting can greatly increase the spindown rate even when the magnetosphere is in a stable steady state. For example, Figures~\ref{fig:rot-early}b and \ref{fig:rot-pseudo-crit}b show stable configurations in which the spindown rate is 1.64 and 1.67 times the spindown rate of the untwisted rotating solution respectively (Figure~\ref{fig:rot-torque-crit}). Then applying the standard dipole spindown relation to the observed spindown rate overestimates the surface magnetic field (i.e. the effective dipole moment is increased by the currents in the twisted magnetosphere).

\subsection{Giant flares}

SGR giant flares are both energetic and rare. The three observed events had brief hard spikes releasing $\sim 10^{44}$--$10^{46}$ erg, and softer pulsating tails radiating $\sim 10^{44}$ erg. The tail emission is believed to come from an electron-positron fireball trapped by the strong magnetic field; since all three flares had similar tail energies, this would suggest that they had similar field strengths, of $10^{14}$--$10^{15}$ G. Therefore in order to explain the $2 \times 10^{46}$ erg spike of SGR~1806-20 \citep[e.g.][]{2005Natur.434.1107P}, one needs to dissipate this much energy using a field with $B \lesssim 10^{15}$~G. 

In the model we have discussed, in which energy is stored in the magnetosphere and released by a magnetospheric instability, the total free energy reservoir is less than the total dipole energy $W_0 = \mu^2/3\rstar^3$, which is $ \sim 3\times 10^{47} $ erg when $B \sim 10^{15}$ G and $\rstar \sim 10^6$ cm. As mentioned above, even a polar cap encompassing half the star's magnetic flux ($\thpc = \pi/4$) can only store $\sim 0.07\, W_0 \sim 2\times 10^{46}$ erg before the onset of instability (assuming that the surface shearing is slow). Only a fraction of this free energy will be converted to radiation by dissipative processes following reconnection, and so it is unlikely that the brightest flare was powered by a sheared polar cap. 

On the other hand, twisting more deeply buried magnetic flux, with footpoints nearer the magnetic equator, allows up to $\sim 0.6\, W_0 \sim 1.8\times 10^{47}$ erg to be stored, enough to power the flares if a significant fraction $\gtrsim 0.1$ of the free energy can be dissipated. We do see such large dissipation fractions in our simulations. This scenario is also consistent with a release of energy close to the star, as may be required to explain the large rotation-frequency pulsations in the soft tails. 

The division into sources that produce giant flares and those that produce smaller bursts may be due to the distribution of shearing on the stellar surface. The more common, less energetic bursts may be caused by shearing at high magnetic latitudes, near the magnetic poles. The localization of twisting near the poles is supported by observations of shrinking hotspots on less active, transient magnetars \citep{2011heep.conf..299B}. For example, the transient AXP XTE J1810-197 experienced an outburst in January 2003 \citep{2004ApJ...609L..21I}. The effective emitting area of a hot spectral component has subsequently decreased \citep{2007ApSS.308...79G}, which can be explained by the shrinking of a current-carrying flux bundle near the magnetic pole \citep{2009ApJ...703.1044B}. This outburst could easily have been powered by the stored free energy of a small twisted polar cap.

This work was supported in part by NASA (NNX-10-AI72G and NNX-10-AN14G), and the DOE (DE-FG02-92-ER40699).

\end{document}